\documentstyle[12pt,psfig]{article}
\makeatletter
\def\@cite#1#2{{[{#1}]\if@tempswa\typeout
{IJCGA warning: optional citation argument
ignored: `#2'} \fi}}

\newcount\@tempcntc
\def\@citex[#1]#2{\if@filesw\immediate\write\@auxout{\string\citation{#2}}\fi
  \@tempcnta\z@\@tempcntb\m@ne\def\@citea{}\@cite{\@for\@citeb:=#2\do
    {\@ifundefined
       {b@\@citeb}{\@citeo\@tempcntb\m@ne\@citea\def\@citea{,}{\bf ?}\@warning
       {Citation `\@citeb' on page \thepage \space undefined}}%
    {\setbox\z@\hbox{\global\@tempcntc0\csname b@\@citeb\endcsname\relax}%
     \ifnum\@tempcntc=\z@ \@citeo\@tempcntb\m@ne
       \@citea\def\@citea{,}\hbox{\csname b@\@citeb\endcsname}%
     \else
      \advance\@tempcntb\@ne
      \ifnum\@tempcntb=\@tempcntc
      \else\advance\@tempcntb\m@ne\@citeo
      \@tempcnta\@tempcntc\@tempcntb\@tempcntc\fi\fi}}\@citeo}{#1}}
\def\@citeo{\ifnum\@tempcnta>\@tempcntb\else\@citea\def\@citea{,}%
  \ifnum\@tempcnta=\@tempcntb\the\@tempcnta\else
   {\advance\@tempcnta\@ne\ifnum\@tempcnta=\@tempcntb \else
\def\@citea{--}\fi
    \advance\@tempcnta\m@ne\the\@tempcnta\@citea\the\@tempcntb}\fi\fi}
\makeatother
\newenvironment{Eqnarray}%
     {\arraycolsep 0.14em\begin{eqnarray}}{\end{eqnarray}}

\def\simlt{\stackrel{<}{{}_\sim}}
\def\simgt{\stackrel{>}{{}_\sim}}
\def\be{\begin{equation}}
\def\ee{\end{equation}}
\def\bear{\be\begin{array}}
\def\eear{\end{array}\ee}
\def\bea{\begin{Eqnarray}}
\def\eea{\end{Eqnarray}}

\def\lsim{\mathrel{\raise.3ex\hbox{$<$\kern-.75em\lower1ex\hbox{$\sim$}}}}
\def\gsim{\mathrel{\raise.3ex\hbox{$>$\kern-.75em\lower1ex\hbox{$\sim$}}}}
\def\ifmath#1{\relax\ifmmode #1\else $#1$\fi}
\def\ls#1{\ifmath{_{\lower1.5pt\hbox{$\scriptstyle #1$}}}}

\def\ie{{\it i.e.}}
\def\eg{{\it e.g.}}

\def\beq{\begin{equation}}
\def\eeq{\end{equation}}
\def\beqa{\begin{Eqnarray}}
\def\eeqa{\end{Eqnarray}}

\def\baselinestretch{1}
\begin{document}
\def\IJMPA #1 #2 #3 {{\sl Int.~J.~Mod.~Phys.}~{\bf A#1}\ (19#2) #3$\,$}
\def\MPLA #1 #2 #3 {{\sl Mod.~Phys.~Lett.}~{\bf A#1}\ (19#2) #3$\,$}
\def\NPB #1 #2 #3 {{\sl Nucl.~Phys.}~{\bf B#1}\ (19#2) #3$\,$}
\def\PLB #1 #2 #3 {{\sl Phys.~Lett.}~{\bf B#1}\ (19#2) #3$\,$}
\def\PR #1 #2 #3 {{\sl Phys.~Rep.}~{\bf#1}\ (19#2) #3$\,$}
\def\JHEP #1 #2 #3 {{\sl JHEP}~{\bf #1}~(19#2)~#3$\,$}
\def\PRD #1 #2 #3 {{\sl Phys.~Rev.}~{\bf D#1}\ (19#2) #3$\,$}
\def\PTP #1 #2 #3 {{\sl Prog.~Theor.~Phys.}~{\bf #1}\ (19#2) #3$\,$}
\def\PRL #1 #2 #3 {{\sl Phys.~Rev.~Lett.}~{\bf#1}\ (19#2) #3$\,$}
\def\RMP #1 #2 #3 {{\sl Rev.~Mod.~Phys.}~{\bf#1}\ (19#2) #3$\,$}
\def\ZPC #1 #2 #3 {{\sl Z.~Phys.}~{\bf C#1}\ (19#2) #3$\,$}
\def\PPNP#1 #2 #3 {{\sl Prog. Part. Nucl. Phys. }{\bf #1} (#2) #3$\,$}

\catcode`@=11
\newtoks\@stequation
\def\subequations{\refstepcounter{equation}%
\edef\@savedequation{\the\c@equation}%
  \@stequation=\expandafter{\theequation}
  \edef\@savedtheequation{\the\@stequation}
  \edef\oldtheequation{\theequation}%
  \setcounter{equation}{0}%
  \def\theequation{\oldtheequation\alph{equation}}}
\def\endsubequations{\setcounter{equation}{\@savedequation}%
  \@stequation=\expandafter{\@savedtheequation}%
  \edef\theequation{\the\@stequation}\global\@ignoretrue

\noindent}
\catcode`@=12
\begin{titlepage}

\title{{\bf General RG Equations\\
 for Physical Neutrino Parameters\\
and their Phenomenological Implications}}
\vskip2in
\author{ 
{\bf J.A. Casas$^{1,2,3}$\footnote{\baselineskip=16pt E-mail: {\tt
casas@mail.cern.ch}}}, 
{\bf J.R. Espinosa$^{2,3,4}$\footnote{\baselineskip=16pt E-mail: {\tt
espinosa@makoki.iem.csic.es}}}, 
{\bf A. Ibarra$^{1}$\footnote{\baselineskip=16pt  E-mail: {\tt
alejandro@makoki.iem.csic.es}}} and 
{\bf I. Navarro$^{1}$\footnote{\baselineskip=16pt E-mail: {\tt
ignacio@makoki.iem.csic.es}}}\\ 
\hspace{3cm}\\
 $^{1}$~{\small I.E.M. (CSIC), Serrano 123, 28006 Madrid, Spain}
\hspace{0.3cm}\\
 $^{2}$~{\small TH-Division, CERN, CH-1211 Geneva 23, Switzerland}
\hspace{0.3cm}\\
 $^{3}$~{\small I.F.T. C-XVI, U.A.M., 28049 Madrid, Spain }
\hspace{0.3cm}\\
 $^{4}$~{\small I.M.A.F.F. (CSIC), Serrano 113 bis, 28006 Madrid, Spain}.
} 
\date{} 
\maketitle 
\def\baselinestretch{1.15} 
\begin{abstract}
\noindent The neutral leptonic sector of the Standard Model presumably
consists of three neutrinos with non-zero Majorana masses with properties
further determined by three mixing
angles and three CP-violating phases. We derive the general
renormalization group equations for these physical parameters and
apply them to study the impact of radiative effects on neutrino
physics. In particular, we examine the
existing solutions to the solar and atmospheric neutrino problems, 
derive conclusions on their theoretical naturalness, and
show how some of the measured neutrino parameters could be 
determined by purely radiative effects.  
For example, the mass
splitting and mixing angle suggested by solar neutrino data could be
entirely explained as a radiative effect if the small angle MSW solution
is realized. On the other hand, the mass
splitting required by atmospheric neutrino data is 
probably determined by unknown physics at a high energy scale.
We also discuss the effect of non-zero CP-violating phases on
radiative corrections.
\end{abstract}

\thispagestyle{empty}
\leftline{CERN-TH/99-315}
\leftline{October 1999}
\leftline{}

\vskip-23cm
\rightline{}
\rightline{IEM-FT-196/99}
\rightline{CERN-TH/99-315}
\rightline{IFT-UAM/CSIC-99-40}
\rightline{hep-ph/9910420}
\vskip3in

\end{titlepage}
\setcounter{footnote}{0} \setcounter{page}{1}
\newpage
\baselineskip=20pt

\noindent

\section{Introduction}

There is mounting experimental evidence \cite{SK,ATM,SOL} that flavour
is not conserved in the flux of atmospheric and solar neutrinos. The most
plausible, simplest and best motivated interpretation for this phenomenon
is that interaction eigenstate neutrinos mix in a non trivial way into
neutrino mass-eigenstates with different non-zero masses, leading to
flavour oscillations \cite{pontecorvo}.  The theoretical scenario that
results most
economical in accounting for the observed neutrino
anomalies\footnote{Leaving aside the, as yet unconfirmed, LSND anomaly
\cite{LSND}.} assumes that the known neutrinos of the Standard Model
($\nu_e,\nu_\mu,\nu_\tau$) acquire Majorana masses through a dimension-5
operator \cite{weinberg}, generated at some high energy scale $\Lambda$
(the model can also be made supersymmetric).

The $3\times 3$ neutrino mass matrix ${\cal M}_\nu$ is diagonalized
according to
\be \label{Udiag}
U^T{\cal M}_\nu U={\mathrm diag}(m_1,m_2,m_3),
\ee
and we can choose $m_i\geq 0$. The MNS \cite{MNS} unitary matrix $U$
relates flavour
and mass eigenstate neutrinos according to $\nu_\alpha=U_{\alpha i}\nu_i$.
The 'CKM' matrix, $V$, is defined through
\be
\label{Udef}
U={\mathrm diag}(e^{i\alpha_e},e^{i\alpha_\mu},e^{i\alpha_\tau})\cdot
V\cdot  
{\mathrm diag}(e^{-i\phi/2},e^{-i\phi'/2},1),  
\ee
and we use the standard parametrization  
\be
V=R_{23}(\theta_1)\cdot {\mathrm
diag}(e^{-i\delta/2},1,e^{i\delta/2})\cdot
R_{31}(\theta_2)\cdot {\mathrm
diag}(e^{i\delta/2},1,e^{-i\delta/2})\cdot
R_{12}(\theta_3),
\ee
where $R_{ij}(\theta_k)$ is a rotation in the $i$-$j$ plane by the mixing 
angle $\theta_k$, that can be taken $0\leq \theta_k\leq \pi/2$ without loss of
generality. Explicitly,
\be \label{Vdef}
V=\pmatrix{c_2c_3 & c_2s_3 & s_2e^{-i\delta}\cr
-c_1s_3-s_1s_2c_3e^{i\delta} & c_1c_3-s_1s_2s_3e^{i\delta} & s_1c_2\cr
s_1s_3-c_1s_2c_3e^{i\delta} & -s_1c_3-c_1s_2s_3e^{i\delta} & c_1c_2\cr}.
\ee
where $s_i\equiv \sin \theta_i$, $c_i\equiv \cos \theta_i$.

For later use, it is convenient to define the intermediate matrix $W$ as
\be
\label{W}
W=V\cdot
{\mathrm diag}(e^{-i\phi/2},e^{-i\phi'/2},1)\ .
\ee
The phases $\alpha_e,\alpha_\mu,\alpha_\tau$, in eq.~(\ref{Udef})
are unphysical and may be
rotated away by a redefinition of the flavour-eigenstate neutrinos
$\nu_e,\nu_\mu,\nu_\tau$, so that $U$ coincides with $W$.
It is useful, however, to keep these phases for the
discussions of the next sections. The phases $\delta$, $\phi$ and $\phi'$
are physical and responsible for CP violation in the leptonic sector
(although only $\delta$ can have an effect in neutrino oscillation
experiments). The physical phases $\phi$ and $\phi'$ (usually
kept out of the diagonalization matrix) are extracted from the equation   
\be V^T{\cal M}_\nu V={\mathrm diag}(\tilde{m}_1,\tilde{m}_2,\tilde{m}_3),
\label{mtildes}
\ee  
where
$\tilde{m}_1=m_1e^{i\phi}$, $\tilde{m}_2=m_2e^{i\phi'}$,
$\tilde{m}_3=m_3$.

Let us now  summarize the experimental information
on the neutrino sector as setting the following bounds:

From SK + CHOOZ data \cite{SK,CHOOZ} we learn that $\theta_2$ is small,
with
\be
\sin^2\theta_2< 0.1 \,\, (0.2),
\label{exp2}
\ee
at $90\%$ ($99\%$) C.L., or  $\sin^22\theta_2<0.36\, (0.64)$. The
smallness of $\theta_2$ implies that the oscillations of atmospheric and
solar neutrinos are dominantly two-flavour oscillations, described by a
single mixing angle $\theta_i$ (precise analysis of the data must also
include the subdominant effects due to a non-zero $\theta_2$
\cite{rujula}). 

In this approximation, oscillations of atmospheric neutrinos are
dominantly $\nu_\mu\rightarrow\nu_\tau$ with \cite{range}
\bea
5\times10^{-4}\ {\rm eV}^2 <  &\Delta m^2_{atm}& < 10^{-2}\ {\rm
eV}^2\ , \nonumber\\ \sin^22\theta_1&>&0.82\ .
\label{atmexp}
\eea

Several parameter choices are possible to interpret the
oscillations of solar neutrinos \cite{SOL}:

The large angle MSW solution (LAMSW), with
\bea 10^{-5}\ {\rm eV}^2 <  &\Delta m^2_{sol}& < 2\times 10^{-4}\ {\rm
eV}^2, \nonumber\\ 0.5 < &\sin^22\theta_3& < 1.
\label{LAMSW}
\eea

The small angle MSW solution (SAMSW), with
\bea 3\times 10^{-6}\ {\rm eV}^2 <  &\Delta m^2_{sol}& < 10^{-5}\
{\rm
eV}^2, \nonumber\\ 2\times 10^{-3} < &\sin^22\theta_3& < 2\times 10^{-2}.
\label{SAMSW}
\eea
        
And the vacuum oscillation solution (VO), with
\bea 5\times10^{-11}\ {\rm eV}^2 <  &\Delta m^2_{sol}& <   
1.1\times10^{-10}\ {\rm eV}^2, \nonumber\\ \sin^22\theta_3 &>& 0.67\, .
\label{VO}
\eea

Other relevant experimental information concerns the non-observation of
neutrinoless double $\beta$-decay, which requires the $ee$ element of the
${\cal M}_\nu$ matrix to be bounded as~\cite{Baudis}
\bea {\cal M}_{ee}
 \equiv
\big\vert  m_{1}\,c_2^2c_3^2e^{i\phi} +
m_{2}\,c_2^2s_3^2e^{i\phi'} + m_{3}\, s_2^2\,e^{i2\delta}
\big\vert\simlt 0.2\ {\rm eV}.
\label{B}
\eea
In addition, Tritium $\beta$-decay
experiments indicate $m_i < 2.5$ eV for any mass eigenstate with
a significant $\nu_e$ component \cite{triti}.
Finally, no experimental information is available yet  on the 
CP-violating phases.

To write the previous bounds, we have followed the standard convention
that
orders the three mass eigenvalues with $m_3$ as the most split eigenvalue,
\ie\ $|\Delta m^2_{21}|<
|\Delta m^2_{32}|, |\Delta m^2_{31}|$  (where $\Delta m^2_{ij}\equiv
m_i^2-m_j^2$) and identifies $|\Delta m^2_{21}|=\Delta m^2_{sol}$,
$|\Delta m^2_{31}|=\Delta m^2_{atm}$.  
Note that observations require $|\Delta m^2_{21}|\ll  |\Delta m^2_{32}|
\sim |\Delta m^2_{31}|$ (see however \cite{CL}).  We do not adopt any
particular ordering for $m_1, m_2$. Nevertheless,  note that although the
interchange of $m_1, m_2$  leaves $\sin^22\theta_1$ and
$\sin^22\theta_2$ unaffected, it  flips the sign of $\cos2\theta_3$,
which is important for the MSW effect \cite{MSW}. 

The relative wealth of experimental data has motivated efforts on the
theoretical
side to understand the possible patterns of neutrino masses and to follow
the hints that such patterns offer on the fundamental symmetries
underlying them. A more modest approach, perhaps more powerful
at this given time, is to study Standard Model (SM) [or
 Minimal Supersymmetric Standard Model (MSSM)] radiative corrections
(of well known origin) to neutrino parameters. Although neutrinos
are famous for having extremely weak couplings to stable matter, this is
not so for all the virtual particles that may appear in loops. Moreover,
as we have seen, some of the mass splittings suggested by the data are
very small and could be easily upset by radiative corrections of modest
size. It turns out that, in many cases of interest, radiative effects have
a very significant impact on neutrino physics. 

Many papers have dealt recently with these issues
\cite{cein1,cein2,cein3,ellis,ellos,BRS,haba,haba2,ma}. The  most
often
considered scenario is to assume that a (Majorana) effective mass
operator for the three light neutrinos is generated by some
unspecified mechanism at a high energy scale, $\Lambda$. If, apart
from this mass operator, the effective theory below $\Lambda$ is just the
SM or the MSSM, the lowest dimension operator of this kind is
\be
\delta{\cal L}=-\frac{1}{4}\kappa_{ij} (H\cdot L_i)(H\cdot L_j) + {\mathrm
h.c.},
\ee
where $H$ is the SM Higgs (or the MSSM Higgs with the 
appropriate  hypercharge), $L_i$ are the lepton doublets,  
and $\kappa_{ij}$ is a (symmetric) matricial coupling.
The neutrino mass matrix is then
${\cal M}_\nu=\kappa v^2/2$, with $v=246$ GeV (with this definition
$\kappa$ and ${\cal M}_{\nu}$ obey the same RGE).
Below the scale $\Lambda$, the
most important radiative corrections to ${\cal
M}_\nu$, which are proportional to $\log(\Lambda/M_Z)$ (where the $Z$ boson 
mass, $M_Z$, sets the low-energy scale of interest), can be most
conveniently computed using renormalization group (RG) techniques. In
short, one has to run ${\cal M}_\nu$ from $\Lambda$ down to $M_Z$ with
the appropriate RGE to
obtain the radiatively corrected neutrino mass matrix. These corrections
have interesting implications for the neutrino masses, the
mixing angles and the CP-violating phases.

As we will see, and has been discussed extensively in the literature,
even if one starts with degenerate neutrino masses at the scale $\Lambda$,
the relevant radiative corrections (with a typical size controlled by
the tau Yukawa coupling squared and a loop suppression factor) tend
 to induce mass splittings 
that can be too large for some of the proposed solutions to the
observed neutrino problems. They are particularly dangerous for the
VO solution to the solar neutrino problem, and just about
right for the SAMSW solution, which is certainly suggestive. They tend to
be too small to explain the atmospheric mass splitting, which presumably
requires to be generated by effects of the physics beyond $\Lambda$. One
interesting possibility that, beginning with degenerate neutrinos, can
produce radiatively a mass pattern in
agreement with experiment is a see-saw scenario, for which the RG
analysis was performed in \cite{cein1,cein2}.

Degenerate (or nearly degenerate) neutrino mass eigenvalues, besides
being a very symmetrical initial condition (as befits the physics  at a
more fundamental scale) offer two important bonuses: the mass splittings
at low-energy would be entirely due to radiative corrections and 
the mixing angles can evolve quickly to some particular values. The reason
for the latter is the following.
In the presence of degenerate mass eigenvalues there is an 
ambiguity in the choice of the eigenvectors which translates to the
definition of the mixing angles. When a perturbation is added that removes
that degeneracy, the ambiguity is resolved in a well defined way (which
depends on the perturbation) and a particular configuration of mixing
angles is chosen. This is exactly what happens when a (sufficiently
precise) initial mass degeneracy of two neutrino masses is lifted by
radiative corrections. In RG language, the mixing angles can be
driven quickly to
infrared quasi-fixed points. Thus, RG effects could in principle be capable
of explaining some of the measured values of the mixing angles.

In this paper we extend previous work of us \cite{cein1,cein2,cein3}
and others \cite{ellis,ellos,BRS,haba,haba2,ma} along these lines.
Assuming 
three flavours of light Majorana neutrinos, we derive the
general renormalization group equations (RGEs) for their three masses,
three mixing angles and three CP-violating phases (information
alternatively encoded in the RGE for the mass eigenvalues and the complex
mixing matrix $U$).  This form of writing the RGEs represents an
advantageous alternative to using the RGE for ${\cal M}_\nu$ and,
after obtaining the radiatively corrected mass matrix, extracting from
it the physical parameters in whatever parametrization it is
chosen. These two alternative methods can be described in short as
'diagonalize and run' versus 'run and diagonalize'. 
One advantage of
the method presented here 
is that  it avoids the proliferation of
unphysical parameters (notice that  ${\cal M}_\nu$ has 12 degrees of
freedom, but only 9 are physical), which allows to keep track of the
physics in a more efficient way. 
Another one 
is that it permits to
write the RGEs in a quite general form, without any reference to a
particular scenario, such as the SM, the MSSM, see-saw, etc. This
allows to appreciate interesting features, \eg\ the mentioned presence
of stable (pseudo infrared fixed-point) directions for mixing angles and
CP phases, which are not  consequence of a particular scenario. 
Finally,
this method allows to determine the physical conditions for the
validity of certain  approximations, such as the two-flavour
approximation, in a reliable way.

In section 2, these general RG formulas are extracted from the most 
general form of the RGE
for the neutrino mass matrix ${\cal M}_\nu$. These can be particularized to
any model of interest. We give as particular examples the
SM, the MSSM and a see-saw scenario. Some conclusions that hold in
general are also discussed. 
In section~3, using the equations derived in the previous section, we
study the implications of radiative corrections for neutrino physics
in the SM case (or the MSSM) neglecting the effect of CP violating
phases, so that the mixing matrix $U$ is real. Section 4 extends this
analysis to the more general case of non-zero CP-violating phases.
Section 5 is devoted to the conclusions and the Appendix contains
some technical details regarding the stability conditions for the mixing
matrix $U$ in the general case, and how they are reached.
 
\section{RGEs for physical parameters}

The energy-scale evolution of the
$3\times 3$ neutrino mass matrix ${\cal M}_\nu$ is generically 
described by a RGE of
the form ($t=\log\mu$): \be \frac{d {\cal M}_\nu}{dt}=-(\kappa_U
{\cal M}_\nu + {\cal M}_\nu P + P^T {\cal M}_\nu), \label{RGM} \ee (in
particular
${\cal M}_\nu^T={\cal M}_\nu$ is preserved). For interesting
frameworks, such as the SM, the MSSM or the  
see-saw mechanism (supersymmetric or not), $\kappa_U$ and $P$ 
are known explicitly \cite{RGSM,babu,cein1}. In
(\ref{RGM}), the term $\kappa_U{\cal M}_\nu$ gives a family-universal
scaling of ${\cal M}_\nu$ which does
not affect its texture, while the non family-universal part (the most
interesting effect) corresponds to the terms that involve the matrix
$P$. 

The goal of this section is to extract the RGEs for the physical neutrino
parameters: the mass eigenvalues, the mixing angles and the CP phases. We
find convenient to work out the RGEs for the completely general case first
(\ie\ without specifying the form of $\kappa_U$ and $P$), and later
specialize them for particular cases of interest.  This allows to
appreciate interesting features, \eg\ the presence of stable (pseudo
infrared fixed-point) directions for angles and phases, which are not a
consequence of the particular framework chosen.
 

Using the parametrization and conventions of sect.~1, 
one gets from eqs.~(\ref{Udiag}, \ref{RGM}), after some amusing algebra, 
the RGEs for the mass eigenvalues and the MNS matrix
\be
\label{RGmass}
\frac{d m_i}{dt}= - 2 m_i \hat P_{ii} - m_i Re(\kappa_U),
\ee
\be
\label{RGU}
\frac{d U}{dt}= U T.
\ee
We have defined
\be 
\label{Pdef}
\hat{P}\equiv \frac{1}{2}U^\dagger (P+P^\dagger) U,
\ee
while $T$ is an anti-hermitian (so that the unitarity of $U$ is preserved
by the RG running) $3\times 3$ matrix with 
\bea
\label{Tgendef}
T_{ii}&\equiv&i \hat{Q}_{ii},\nonumber\vspace{.5cm}\\
T_{ij}&\equiv&\frac{1}{(m_i^2-m_j^2)}\left[(m_i^2+m_j^2)\hat{P}_{ij}
+2 m_i m_j \hat{P}_{ij}^*\right]+i\hat{Q}_{ij}\nonumber\vspace{.5cm}\\
&=&\nabla_{ij} Re (\hat P_{ij})+i[\nabla_{ij}]^{-1} Im
(\hat P_{ij})+i\hat{Q}_{ij},\hspace{1cm} i\neq j,
\eea
where
\be
\hat{Q}\equiv -\frac{i}{2}U^\dagger (P-P^\dagger) U,
\ee
and
\be
\label{deltadef}
\nabla_{ij}\equiv\frac{m_i+m_j}{m_i-m_j}.
\ee
With the above definitions we can write
\be
U^\dagger P U = \hat{P} + i \hat{Q},
\ee
with $\hat{P}$ and $\hat{Q}$ hermitian. Note that the RGE for $U$ does
not depend on the universal factor $\kappa_U$, as expected.

From eqs.~(\ref{RGU}--\ref{deltadef}), we can derive the 
general RGEs for the mixing angles, the CP-violating phases and the 
unphysical phases.
For this task, it is useful
to define 
\be
\tilde T_{21}=T_{21}e^{i(\phi-\phi')/2},\;\;\;
\tilde T_{31}=T_{31} e^{i\phi/2},\;\;\;
\tilde T_{32}=T_{32}e^{i\phi'/2},
\ee
which allows to conveniently absorb\footnote{The formulas we have written
do not depend on our convention $m_i\geq 0$, which is just a definition for
$\phi$ and $\phi'$, and we could change at will the sign of $m_i$. The
definition of the quantities $\tilde{T}_{ij}$ makes them independent of
such choices of sign conventions.} the phases $\phi$ and $\phi'$
everywhere.
Then we find the following expressions for the RGEs for the mixing angles:
\bea  
\label{RGt1}
\frac{d \theta_1}{dt}=
\frac{1}{c_2} && Re \left[
s_3 \tilde T_{31}  -
c_3 \tilde T_{32} \right],
\eea
\bea
\label{RGt2}
\frac{d \theta_2}{dt}=
 &&-Re \left[
\left(c_3 \tilde T_{31} +
s_3 \tilde T_{32}\right) e^{-i\delta}\right],
\eea
\bea
\label{RGt3}
\frac{d \theta_3}{dt}=
-\frac{1}{c_2} &&Re \left[
c_2 \tilde T_{21} +
\left( s_3 \tilde T_{31} -
 c_3 \tilde T_{32}\right)s_2 e^{-i\delta}\right];
\eea
the RGEs for the CP-violating phases:
\be
\label{RGCP1}
\frac{d \delta}{dt}= Im \left[
\frac{1}{s_3c_3} \tilde T_{21}
+\left(\frac{V_{21} }{c_1c_2s_2}e^{-i\delta}
-\frac{c_3V^*_{21}}{s_1c_2s_3}
\right)\tilde T_{32}-
 \left(\frac{s_3V_{22}^* }{s_1c_2c_3}
+ \frac{V_{22}}{c_1c_2s_2}e^{-i\delta}   
\right)\tilde T_{31}
\right], 
\ee 
\be
\label{RGCP2}
\frac{1}{2}\frac{d \phi}{dt}=   Im  \left[ 
\frac{c_3}{s_3} \tilde T_{21}
+\left(
\frac{V^*_{32}}{c_1c_2}-\frac{c_3V^*_{21}}{s_1c_2s_3}
\right)\tilde T_{32}
+ \left(\frac{V_{31}^*}{c_1c_2}
+ \frac{V_{21}^*}{s_1c_2}\right)\tilde T_{31} 
\right]+\hat{Q}_{33}-\hat{Q}_{11},
\ee
\be
\label{RGCP3}
\frac{1}{2}\frac{d \phi'}{dt}=Im \left[
\frac{s_3}{c_3} \tilde T_{21} +
\left(\frac{V_{32}^*}{c_1c_2}+  
\frac{V^*_{22}}{s_1c_2}
\right)\tilde T_{32}
+ \left(\frac{V_{31}^*}{c_1c_2}
- \frac{s_3V_{22}^*}{s_1c_2c_3}\right)\tilde T_{31}
\right]+\hat{Q}_{33}-\hat{Q}_{22};
\ee
and the RGEs for the unphysical phases:
\be\label{alfae}
\frac{d \alpha_e}{dt}=  Im  \left[
\frac{1}{c_3s_3} \tilde T_{21}
+\left(
\frac{V^*_{32}}{c_1c_2}-\frac{c_3V^*_{21}}{s_1c_2s_3}
\right)\tilde T_{32}
+ \left(\frac{V_{31}^*}{c_1c_2}
-\frac{s_3V_{22}^*}{s_1c_2c_3}\right)\tilde T_{31}
\right]+\hat{Q}_{33},
\ee
\be
\frac{d \alpha_\mu}{dt}=
\frac{1}{s_1c_2} Im \left[
V_{21}^*\tilde T_{31}  +
V_{22}^*\tilde T_{32}\right]+\hat{Q}_{33},
\ee
\be
\frac{d \alpha_\tau}{dt}=
\frac{1}{c_1c_2} Im \left[
V_{31}^*\tilde T_{31}  +
V_{32}^*\tilde T_{32}\right]+\hat{Q}_{33}\ .
\label{alfatau}
\ee

Some comments are in order at this point: 1) the RGE (\ref{RGU}) for $U$
is only satisfied if the unphysical phases are included [that is, in the
general case $W$ does not satisfy eq.~(\ref{RGU})]. The reason for this is 
that the phases $\alpha_{e},\alpha_{\mu},\alpha_{\tau}$, that translate
between $U$ and the physical mixing matrix $W$, depend on the details in
${\cal M}_\nu$. When these details change through RG running the phases
to be rotated away also change [thus eqs.~(\ref{alfae}--\ref{alfatau})].
However,
the
explicit RGEs for the physical parameters $m_i$ and $\delta,\phi,\phi'$ do
not depend on $\alpha_e,\alpha_\mu,\alpha_\tau$ as it should be the case;
2) from (\ref{RGCP1}--\ref{RGCP3}) we also see that, if no phases are
present in $U$ originally, they are not generated by radiative
corrections (unless $P$ contains phases. If it does, one could generate
radiatively small non-zero phases in $U$); 
3) from the structure of $T_{ij}$ (or $\nabla_{ij}$) we see
that large renormalization effects can be expected in two cases: a) if
some couplings in $P$ are large (not the case of the SM) or b) if there
are (quasi)-degenerate mass eigenstates at tree level (causing
$|\nabla_{ij}|\gg 1$).

An aspect of central importance in the discussion  of the RG effects is
that when the neutrino spectrum has (quasi)-degenerate eigenvalues,
one expects large
(even infinite for exact degeneracy) contributions to $dU/dt$.  The
reason is the following. When two mass eigenvalues, say $i$ and $j$,
are equal, there is an ambiguity in the choice of the associated
eigenvectors, and thus in the definition of $U$.  In particular, the
columns $i,j$ could
be rotated at will, \ie\ the matrix $UR_{ij}$, where $R$ is an arbitrary 
 rotation in
the $i$-$j$ plane, will also diagonalize the initial ${\cal M}_\nu$
matrix.  When the perturbation due to RG running is added, the
degeneracy $m_i=m_j$ will be normally lifted  and a particular
rotation of the columns $i,j$ will be singled out: $U'=UR_{ij}$, giving
 $T'_{ij}\simeq 0$ and  removing the singularity in eq.~(\ref{RGU}).
The form of $R$ is thus determined by the matrix $P$. In the 
Appendix we give the derivation of $R$ for a general $P$. 
The matrix $U'$ may be very different from the form of $U$ originally
chosen, thus the initial big jump in the evolution of $U$.
It is easy to see, however, that the subsequent  
evolution of the matrix $U$ will be smooth,
even in the presence of a large $|\nabla_{ij}|$.  If the initial
degeneracy is not exact, but nearly so, then
$|\nabla_{ij}|\gg 1$
and the initial $U$, whatever it is,   will be rapidly driven by
the RG-running close to  the stable $U'$ form.  Thus, $U'$ plays the
role of an infrared pseudo-fixed point. 
Therefore, if some initial neutrino masses are (exactly or
approximately)  degenerate, we have the interesting possibility of
predicting some low-energy  mixing angles or CP phases just from radiative
corrections. This possibility will be exploited in the next sections.

It is also useful to indicate that, in cases for which one particular 
$\tilde{T}_{ij}$ gives the dominant contribution to the RGEs, the
following quantities are approximately conserved (here $k$ can take any
value):
\bea
\label{CONS}
&|U_{ki}|^2+|U_{kj}|^2&,\\
&Im(U_{ki}^*U_{kj})&,\label{CONS2}\\
&[\nabla_{ij}]^{-1}Re(\hat{P}_{ij})&,
\label{CONS3}
\eea
as can re readily proved by using (\ref{RGU}) and keeping only terms
$\sim T_{ij}$ (for the last conserved quantity we assume that $dP/dt$
is never
of the order of $T_{ij}$, which is quite reasonable to expect).

We turn now to particularly interesting scenarios.
In the SM the expressions for $\kappa_U$ and $P$ are given by
\cite{RGSM,babu} 
\be 
\label{KSM}
\kappa_U=\frac{1}{16\pi^2}[3g_2^2-2\lambda-6h_t^2-2{\mathrm Tr}({\bf
Y_e^\dagger
Y_e})], \ee 
where $g_2,\lambda, h_t, {\bf Y_e}$ are the $SU(2)$ gauge
coupling, the quartic Higgs coupling, the top-Yukawa coupling and the
matrix of Yukawa couplings for the charged leptons, respectively; and 
\be
\label{SMP} P=\frac{1}{32\pi^2}{\bf Y_e^\dagger Y_e}\simeq
\frac{h_\tau^2}{32\pi^2}{\mathrm diag}(0,0,1), 
\ee 
where $h_\tau$ is the tau-Yukawa coupling. In the following, we will work
in the approximation of neglecting the electron and muon Yukawa couplings.

In the MSSM one has instead \cite{RGSM,babu}
\be \label{KMSSM}
\kappa_U=\frac{1}{16\pi^2}\left[
\frac{6}{5}g_1^2+6g_2^2-6\frac{h_t^2}{\sin^2\beta}\right],
\ee where $g_1$ is the $U(1)$ gauge coupling, $\tan\beta$ is the
supersymmetric parameter given by the ratio of the two Higgs vevs; 
and 
\be
\label{MSSMP} P=-\frac{1}{16\pi^2}\frac{\bf Y_e^\dagger
Y_e}{\cos^2\beta}\simeq
-\frac{1}{16\pi^2}\frac{h_\tau^2}{\cos^2\beta}{\mathrm diag}(0,0,1). 
\ee
Note that we could have written these equations in terms of the
supersymmetric couplings $\tilde{h}_t=h_t/sin\beta$,
$\tilde{h}_\tau=h_\tau/\cos\beta$ and ${\bf \tilde{Y}_e}={\bf
Y_e}/\cos\beta$.
 
In both cases, the previous formulas (\ref{RGmass}, \ref{RGU})
apply with [see (\ref{SMP}--\ref{MSSMP})] 
\bea
\label{PSMdef}
\hat P_{ii} &=& -\kappa_\tau |U_{3i}|^2,
\vspace{0.1cm}\\
T_{ii}&\equiv&i\hat{Q}_{ii}=0,
\vspace{0.1cm}\\
\label{TSMdef} 
T_{ij}&=&\kappa_\tau\left[\nabla_{ij}Re
(U_{3i}^*U_{3j})+i\nabla_{ij}^{-1}Im (U_{3i}^*U_{3j})\right],
\eea 
where
\be\label{kappaSM}
\kappa_\tau=\frac{h_\tau^2}{32 \pi^2}, \ee for the SM case, and \be
\label{kappaMSSM}\kappa_\tau=-\frac{1}{16
\pi^2}\frac{h_\tau^2}{\cos^2\beta}, 
\ee
for the MSSM (note how $\kappa_\tau$ is enhanced for large $\tan\beta$).
For future use we also define the related quantity 
\be
\epsilon_\tau\equiv\kappa_\tau\log(\Lambda/M_Z).
\ee
Hence, the RGEs for the neutrino masses given by
eq.~(\ref{RGmass}), as well as the RGEs for the mixing angles, the
CP-violating
phases and the unphysical phases given by eqs.~(\ref{RGt1}--\ref{alfatau}),
hold with the substitutions of eqs.~(\ref{PSMdef}, \ref{TSMdef}).

One can apply the general results to study
other cases of interest beyond the SM or the MSSM. One particularly
relevant case is the see-saw scenario \cite{seesaw}. The model includes
three heavy right-handed neutrinos $N_i$ with
Majorana masses given by a $3\times 3$ matrix ${\cal M}$ with overall scale
$M\gg M_Z$ such that the see-saw mechanism is implemented. Above $M$, the
effective neutrino mass matrix is given by 
\be \label{Mseesaw} {\cal
M}_\nu=m_D^T{\cal M}^{-1}m_D, 
\ee 
where $m_D$ is an ordinary Dirac mass matrix coming
from the conventional Yukawa couplings between the left-handed neutrinos,
$\nu_e,\nu_\mu,\nu_\tau$, and the right-handed ones.
The running of ${\cal M}$ and $m_D$ is of the general form \be
\label{RGMM} \frac{d {\cal M}}{dt}={\cal M} P_M + P_M^T {\cal M}, \ee and
\be \label{RGMD} \frac{d m_D}{dt}= m_D(\kappa_U' I_3 + P_D), \ee where
$\kappa_U'$ gives a family universal contribution. Combining (\ref{RGMM})
and (\ref{RGMD}) with (\ref{Mseesaw}) one gets an RGE for ${\cal M}_\nu$
above the Majorana scale $M$ of the form (\ref{RGM}) with 
\be
P=m_D^{-1}P_M m_D-P_D,
\ee 
where we have assumed that $m_D$ has an inverse.
Applying the general formulas to this
particular case one can then easily reproduce previous results derived in
the
literature for this kind of scenarios \cite{cein1,cein2}.

\section{Implications (real case)}

The case in which CP-violating phases are zero, and thus the 
'CKM' matrix $V$ is real, is the one most extensively
studied in the literature, perhaps because there is not any
information yet about leptonic CP violation. 
This section is devoted to the study of this particularly 
interesting scenario, assuming that 
below the high-energy scale, $\Lambda$,
the effective theory is just the SM or the MSSM, with an effective 
Majorana mass matrix for the neutrinos. It is convenient 
in this real case to work with the mass eigenvalues $\tilde m_i$
defined in eq.~(\ref{mtildes})
\be
\tilde{m}_1=m_1e^{i\phi},\hspace{1cm}\tilde{m}_2=m_2e^{i\phi'},
\hspace{1cm}\tilde{m}_3=m_3,
\label{mtildes2}
\ee
where $\phi, \phi'= 0$ or $\pi$.
All $\tilde m_i$'s are now real, but $\tilde m_1$ and
$\tilde m_2$ can be negative.
We also define the quantities $\tilde \nabla_{ij}$ as
\be
\label{nablatilde}
\tilde \nabla_{ij}\equiv\frac{\tilde m_i+\tilde m_j}{\tilde m_i-\tilde m_j}.
\ee
Then, neglecting all the charged-lepton Yukawa couplings 
except $h_\tau$, the RG equations for 
the mass eigenvalues and the matrix $V$ read
\be
\frac{d\tilde m_i}{dt}= -2 \kappa_\tau\tilde  m_i V_{3i}^2 - \tilde m_i
\kappa_U, 
\label{miRG}
\ee 
\bea 
\frac{d V}{dt} =  V T,
\label{URG}
\eea
where $\kappa_U$ is given in eqs.~(\ref{KSM}, \ref{KMSSM}),
$\kappa_\tau$ is given in eqs.~(\ref{kappaSM},\ref{kappaMSSM}) and
\bea
T_{ii}&=&0,\vspace{0.1cm}\nonumber\\
\label{Treal} 
T_{ij}&=&\kappa_\tau\tilde \nabla_{ij}V_{3i}V_{3j}\,\,\, (i\neq j)\  .
\eea
The RGEs for the mixing angles are 
\be  
\label{teta1}\frac{d \theta_1}{dt}=-\kappa_\tau c_1 ( -s_3 V_{31} \tilde
\nabla_{31}+ c_3 V_{32} \tilde \nabla_{32}),  \ee  
\be 
\frac{d
\theta_2}{dt}=-\kappa_\tau c_1 c_2 ( c_3 V_{31} \tilde \nabla_{31}+
s_3 V_{32} \tilde \nabla_{32}),  
\ee  
\be 
\label{teta3}
\frac{d \theta_3}{dt}=-\kappa_\tau
( c_1 s_2 s_3 V_{31} \tilde \nabla_{31}- c_1 s_2 c_3 V_{32} \tilde
\nabla_{32} +V_{31}V_{32}\tilde \nabla_{21})\ .  
\ee  

From eqs.~(\ref{URG}, \ref{Treal}), the
first conclusion is that the radiative corrections to
the  matrix $V$ will be very small unless $|\kappa_\tau\tilde
\nabla_{ij}| \simgt 1$ for some $i,j$. This generically requires
mass degeneracy (both in the absolute value and the sign),
except for the supersymmetric case with very large
$\tan\beta$, and thus large $\kappa_\tau$. The interesting cases
then are a completely degenerate
spectrum $m_1^2\simeq m_2^2\simeq m_3^2$ or an intermediate one with
$m_1^2\simeq
m_2^2\not\simeq m_3^2$.

The initial degree of degeneracy of two given neutrino masses, say $i,j$,
plays a crucial role for the subsequent RG evolution of $V$. In particular, it
is important how small the initial mass splitting 
$\Delta \tilde m_{ij}^{(0)}$ is
compared with the typical one generated by RG evolution $|\delta_{RGE}\Delta
\tilde m_{ij}|\sim |\epsilon_\tau| m_0$, where $m_0$ is 
the average mass scale of the
two neutrinos. There are four possibilities: 1) 
If $\Delta \tilde m_{ij}^{(0)}=0$, 
there is exact degeneracy and $V$ jumps
immediately to $V'$ (note that we could have redefined the initial $V$ to
coincide exactly with the stable matrix $V'$ to start with, and then the
evolution is smooth). 2) If $|\Delta \tilde m_{ij}^{(0)}|\simlt 
2|\kappa_\tau| m_0$ (\ie\ $|\kappa_\tau\tilde \nabla_{ij}|\simgt 1$), 
$V$ will quickly reach $V'$ (we will
be more precise in the next subsection). 3) If
$2|\kappa_\tau| m_0\simlt |\Delta \tilde m_{ij}^{(0)}|
\simlt 2|\epsilon_\tau | m_0$
(\ie\ , $[\log(\Lambda/M_Z)]^{-1}\simlt|\kappa_\tau\tilde 
\nabla_{ij}|\simlt 1$),
$V$ will tend to approach $V'$ but may not reach it
if the total amount of running is not long enough (other possible effects
in this case will be discussed later on). 4) Finally, if 
$|\Delta \tilde m_{ij}^{(0)}|\simgt
2|\epsilon_\tau | m_0$, no dramatic effects are expected from RG
corrections to $V$ (even if still the masses can be considered 
degenerate in the sense $|\Delta \tilde m_{ij}^{(0)}|/m_0\ll 1$).

For the discussion of cases of physical relevance, we will assume that  the
solar neutrino problem is solved by one of the standard solutions, so that
$|\Delta m^2_{21}|\ll  |\Delta m^2_{32}| \sim |\Delta m^2_{31}|$.
In some of the cases that we will analyze (in fact, in the most
interesting cases) the initial values for some of the $m_i$'s are
degenerate, so there exists an ambiguity in the labelling, which
generically is removed after RG running. If the initial $m_i$'s are
only approximately degenerate (which we call quasi-degenerate), in
principle there is no such ambiguity, but the 
RG running may alter the relative size of the $\Delta m^2_{ij}$ splittings, 
thus requiring a relabelling of the $m_3$ eigenvalue at low energy.
Let us study in turn the impact (and physical implications) of the RG
running on the neutrino masses and the mixing angles.

\subsection{Mixing angles}

As we have discussed, in cases with some (sufficiently) degenerate neutrino
masses, large  RG corrections to the matrix $U$ have
the effect of driving it close to a stable form (an infrared pseudo-fixed
point). This rises the interesting possibility of predicting (at least
partially) low-energy  mixing angles just from radiative effects.
In the event of $m_i\simeq m_j$, the stable form of $V$,
say $V'$, is characterized by the condition $T'_{ij}\simeq 0$, which
removes
potential singularities in eq.~(\ref{URG}), and requires
$V'_{3i}=0$ or $V'_{3j}=0$. The sign
of the initial $\kappa_\tau\tilde \nabla_{ij}$ will determine which
one is realized, as we will see shortly. 

We can estimate more precisely how $V$ evolves to $V'$ in the
following way. Suppose there are two nearly degenerate neutrinos with 
$|\epsilon_\tau\tilde \nabla_{ij}|\simgt 1$, so that this term dominates
the RGEs of $V_{3i}$, $V_{3j}$ and $\tilde \nabla_{ij}$ itself:
\bea 
\frac{d V_{3i}^2}{dt} \simeq -\frac{d V_{3j}^2}{dt}=
-2\kappa_\tau V_{3i}^2 V_{3j}^2 \tilde \nabla_{ij}\ ,
\label{V3ij}
\eea
\bea 
\frac{d \tilde \nabla_{ij}^{-1}}{dt} \simeq \kappa_\tau \left(V_{3j}^2 -
V_{3i}^2\right) .
\label{nablaRG}
\eea
Eq.~(\ref{V3ij}) implies that, in the infrared, $V_{3j}\rightarrow 0$
($V_{3i}\rightarrow
0$) for positive (negative) initial $\kappa_\tau\tilde \nabla_{ij}$.
Without loss of generality, we can choose here the
labels $i,j$ so that $\kappa_\tau\tilde \nabla_{ij}^{(0)}>0$, 
where the superscript $(0)$ denotes the initial (high-energy) value.
Hence, $V_{3j}\rightarrow 0$. 
Eqs.~(\ref{V3ij}, \ref{nablaRG}) imply
\bea 
\frac{d}{dt} \left[\frac{V_{3i}V_{3j}}{\tilde \nabla_{ij}}\right]\simeq 0 .
\label{V3ijnabla}
\eea
The conserved quantity $V_{3i}V_{3j}/\tilde \nabla_{ij}$ is the SM real
version of the general conserved quantity given by eq.~(\ref{CONS3}).
Notice that $[\tilde \nabla_{ij}]^{-1}=(\tilde m_i^2-\tilde m_j^2)/
(\tilde m_i+\tilde m_j)^2$ is a measure of the relative splitting of 
$\tilde m_i$ and $\tilde m_j$. For quasi-degenerate masses,
$[\tilde \nabla_{ij}]^{-1}\simeq \Delta m^2_{ij}/4m^2$.

From (\ref{nablaRG}) we find
\be
[\nabla_{ij}]^{-1}=[\nabla_{ij}^{(0)}]^{-1}-
\epsilon_\tau\langle V_{3j}^2-V_{3i}^2\rangle,
\label{nablainteg}
\ee
where by $\langle V_{3j}^2-V_{3i}^2\rangle$ we mean the average value in
the interval of running. When $V\rightarrow V'$ quickly, we can estimate
that, with our choice of indexes
\be
\label{average}
\langle V_{3j}^2-V_{3i}^2\rangle\sim
({V'}_{3j}^2-{V'}_{3i}^2)\sim -\frac{1}{2},
\ee
(note that $V'_{3j}\rightarrow 0$ by the running, while
${V'}_{3i}^2\sim 1/2$ is required to
agree with experimental indications).

Using (\ref{nablainteg}) in the conservation equation (\ref{V3ijnabla}) we
end up with the useful relation
\be
\frac{ V_{3i} V_{3j} }{V^{(0)}_{3i} V^{(0)}_{3j}}=
\left[ 1- \epsilon_\tau \langle
V_{3j}^2-V_{3i}^2 \rangle \tilde \nabla_{ij}^{(0)} \right]^{-1}.
\label{deltaRGE3}
\ee
This equation is interesting since it gives a measure of the change in
$V_{3i}V_{3j}$, and thus in $V$, as a function of the initial $\tilde
\nabla_{ij}$ and the amount of running. The stable form of $V$ is
achieved for $V_{3i}V_{3j}\rightarrow 0$. From
eqs.~(\ref{average},\ref{deltaRGE3}) we
can evaluate the initial $\tilde \nabla_{ij}$ to get
$V_{3i}V_{3j}= V_{3i}^{(0)}V_{3j}^{(0)}/F$, where $F$ is an arbitrary
factor
\bea 
V_{3i}V_{3j}= V_{3i}^{(0)}V_{3j}^{(0)}/F\  \Rightarrow \ 
\kappa_\tau\tilde
\nabla_{ij}^{(0)}\simeq\frac{2(F-1)}{\log\frac{\Lambda}{M_Z}}.
\label{arrows}
\eea
If $V$ finishes close to the stable form, that means $F\gg 1$.
For example, for $F> 10$ and $\Lambda=10^{10}$ GeV we need 
$\kappa_\tau\tilde \nabla_{ij}^{(0)}>1$. 

Notice that the condition to get $F\gg 1$, from eq.~(\ref{arrows}),
can be restated as $[\Delta m_{ij}^2]^{(0)}\ll
2\epsilon_\tau m_0^2\sim \delta_{RGE}\Delta m_{ij}^2$.
This means that the final
value for $\Delta m_{ij}^2$ must be essentially given 
by $\delta_{RGE}\Delta m_{ij}^2$, that is
\bea
\Delta m^2_{ij}\simeq 2\kappa_\tau m^2\log\frac{\Lambda}{M_Z} . 
\label{deltaRGE5}
\eea
This is interesting since it implies that
if some $|\tilde
\nabla_{ij}|$ is initially large enough to drive the matrix
$V$ into a stable form, not only the final mixing angles, but also the 
final $\Delta m^2_{ij}$ splitting will be determined just 
by the radiative corrections. This result will be useful for 
the analysis of cases of physical interest.

If $V$ does not finish close to the stable form, $|V_{3i}V_{3j}|$ may
increase instead of going to zero (thus the value of $F$ could 
be less than 1). For this to happen, one must start with
$\kappa_\tau V_{3i}^2<\kappa_\tau V_{3j}^2$. Then 
$[\nabla_{ij}]^{-1}$ decreases (as the mass
eigenvalues get closer) and by eq.~(\ref{V3ijnabla}), $|V_{3i}V_{3j}|$
initially grows. If the
running stops before the eventual decreasing of $|V_{3i}V_{3j}|$ below its 
initial value, then $\epsilon_\tau\langle V_{3j}^2-V_{3i}^2\rangle >0$ in
eq.~(\ref{deltaRGE3}). Notice that a large increasing in 
$V_{3i}V_{3j}$ needs a cancellation in that equation,
namely $\frac{1}{2}|\kappa_\tau| \log\frac{\Lambda}{M_Z}\simeq
(1-F)\left|\tilde \nabla_{ij}^{(0)}\right|^{-1}$, which implies a 
certain amount of fine-tuning (the initial mass splitting must be of the
order of the one generated by running, which implies a correlation
between two quantities of totally different physical origins). 

In the rest of the section we will consider the physics of the 
different types of spectrum which are relevant for the evolution 
of $V$ (and thus of the mixing angles). They are the following:
\bea
{\em (i)}&&\hspace{0.7cm} 
\tilde m_a \simeq \tilde m_b\simeq\tilde m_c 
\Rightarrow  |\tilde \nabla_{ab}|, |\tilde \nabla_{bc}|, 
                                      |\tilde \nabla_{ac}| \gg 1,
\nonumber\\
{\em (ii)}&&\hspace{0.7cm} 
-\tilde m_a \simeq -\tilde m_b\simeq 
\tilde m_c \Rightarrow  |\tilde \nabla_{ab}|\gg 1,
\nonumber\\
{\em (iii)}&&\hspace{0.7cm} 
\tilde m_a \simeq \tilde m_b\not \simeq \pm\tilde m_c 
\Rightarrow  |\tilde \nabla_{ab}|\gg 1,
\nonumber\\
{\em (iv)}&&\hspace{0.7cm} 
-\tilde m_a \simeq \tilde m_b\not\simeq \pm \tilde m_c .
\label{cases}
\eea
We leave the indices unspecified as they will be determined only after RG
evolution (recall that $m_3$ is defined as the most split eigenvalue).
From the point of view of the running of the matrix $V$, what
matters are the  sizable $|\tilde \nabla_{ij}|$. So, case {\em (iv)} is
trivial (the matrix $V$ changes little) and cases {\em (ii)}--{\em (iii)}
can be analyzed simultaneously.

Before embarking in the discussion of these cases, note that experimental  
observations require 
\be \label{final}
V_{3i}\simeq (s_3/\sqrt{2},-c_3/\sqrt{2},\pm
1/\sqrt{2}),
\ee
with different values for $\theta_3$ depending on the choice of solution
to the solar neutrino problem.

\vspace{0.3cm}
\noindent
{\em Case (i)}

\vspace{0.2cm}

\noindent
In case {\em (i)}, if $|\kappa_\tau\tilde \nabla_{ij}|> 1$  for all 
$i,j$ pairs, then $V\rightarrow V'$ with $V'_{3i}V'_{3j}\rightarrow 0$, 
and this is obviously incompatible with the desired form (\ref{final}).
Possible ways out would require starting with nearly degenerate 
masses but not so close as to drive $V$ to its infrared fixed-point form.
Such scenarios really belong in one of the  next cases.

\vspace{0.3cm}
\noindent
{\em Cases (ii)-- (iii)}

\vspace{0.2cm}

\noindent
Since $m_3$ is by definition the most split eigenvalue, it is clear that
in case {\em (iii)} the label $c$ corresponds to 3. This may not be so 
for case {\em (ii)}, although we will see that a correct 
phenomenology requires it too. For the moment we will maintain the 
$a,b,c$ labels.

In these cases the RGEs for the matrix
$V$ are dominated by the $T_{ab}$
terms. Explicitly,
\bea 
\frac{d V_{3a}}{dt} &\simeq& -\kappa_\tau V_{3b}^2 V_{3a} \tilde
\nabla_{ab},
\nonumber\\
\frac{d V_{3b}}{dt} &\simeq& \kappa_\tau V_{3a}^2 V_{3b} \tilde
\nabla_{ab},
\nonumber\\
\frac{d V_{3c}}{dt} &\simeq& 0.
\label{U3iRG}
\eea
If initially $\kappa_\tau\tilde \nabla_{ab}>0$ ($\kappa_\tau\tilde
\nabla_{ab}<0$), then $V_{3b}\rightarrow 0$ ($V_{3a}\rightarrow 0$)
in the infrared. From (\ref{final}) we see that only $V_{31}$ or 
$V_{32}$ can be accepted to vanish [actually, this is guaranteed for case
{\em (iii)}, since $c=3$]. As we are free to interchange the labels 1,2,
we choose $V_{31}\rightarrow 1$. 
On the other hand, the radiatively corrected
masses are given by the expression (we absorb the universal scaling factor
in a multiplicative redefinition of $\tilde m_i$)
\be
\label{shifts}
\tilde m_i\rightarrow \tilde m_i (1+2\epsilon_\tau\langle V_{3i}^2\rangle),
\ee
where, as usual, $\langle V_{3i}^2\rangle$ is the average value of
$V_{3i}^2$ in the interval of running. If $|\kappa_\tau
\tilde\nabla_{ij}|>1$ then $V\rightarrow V'$ quickly and we can set
$\langle V_{3i}^2\rangle\simeq {V'}_{3i}^2$. 

Eq.~(\ref{final}) and $V_{31}\rightarrow 0$ imply that ${V'}_{32}^2\simeq
{V'}_{33}^2\simeq 1/2$. We have then, from eq.~(\ref{shifts}),
\bea
\tilde m_1&\rightarrow & \tilde m_1,\vspace{0.1cm}\nonumber\\
\tilde m_2&\rightarrow & \tilde
m_2(1+\epsilon_\tau),\vspace{0.1cm}\nonumber\\
\tilde m_3&\rightarrow &\tilde m_3(1+\epsilon_\tau).\vspace{0.1cm}
\label{123}
\eea 
The conventional labelling requires $|\Delta m_{32}^2|$, $|\Delta
m_{31}^2|$ $\gg|\Delta m_{21}^2|$, so $\tilde m_2^2$ and $\tilde m_3^2$ must
have an initial splitting large enough $\sim \Delta m_{atm}^2$ from the
beginning (it will not be generated by the running)\footnote{Conversely,
if we do not start with such a splitting, eq.~(\ref{123}) indicates that
$m_1$ becomes the most 
split eigenvalue, so it should be relabelled as $m_3$, leading to 
$V_{33}\rightarrow 0$, which is not acceptable.}.
From  eq.~(\ref{cases}) 
we conclude that also in case {\em (ii)} $c$ must be 3, 
in order to guarantee a correct phenomenology.
On the other hand, as was discussed before,
if initially $|\kappa_\tau \tilde \nabla_{21}|> 1$, 
so that $V$ reaches the stable form $V'$, then 
the low-energy mass splitting is
essentially the one generated by the running:
\be
\Delta m_{21}^2\simeq 2|\epsilon_\tau| m_0^2.
\ee
Interestingly enough, this can be naturally
of the right size for the SAMSW solution 
($3\times10^{-6}\ {\rm eV}^2 < 
\Delta m^2_{sol} < 10^{-5}\ {\rm eV}^2$). For example, in the intermediate 
cases {\em (iii)} this is achieved with sizeable values of the 
cut-off ($\Lambda\simgt 10^{12}\ {\rm GeV}$) and/or working 
in the supersymmetric scenario.

The final (stable) form of $V$ in this case is
\bea 
V'\simeq \pmatrix{V_{11}' &   V_{12}'  &   V_{13}^{(0)} \cr
V_{21}' &   V_{22}'  &  V_{23}^{(0)} \cr
0 & V_{32}' & V_{33}^{(0)}  \cr}\ ,
\label{Ustable2}
\eea 
where $V_{ij}^{(0)}$ denote the initial $V_{ij}$ values. The values 
of the remaining $V_{ij}'$ can be straightforwardly written in terms 
of the former using the unitarity of the matrix $V'$ and the condition
$V'_{31}=0$. Moreover, 
we know that at low energy $s_2\simeq 0$, $s_1\simeq \pm 1/\sqrt{2}$,
which allows to fill the third column of eq.~(\ref{Ustable2}), 
and thus the complete final matrix $V'$
\bea 
V'\simeq \pmatrix{1 &   0  &  0 \cr
0 &    1/\sqrt{2} &  1/\sqrt{2}  \cr
0 &  \mp 1/\sqrt{2} & \pm 1/\sqrt{2}  \cr}\ ,
\label{Ustable22}
\eea 
This implies $\sin^2 2\theta_3\simeq 0$, which is also consistent with
the SAMSW solution. It is worth-stressing that the value of 
$\sin^2 2\theta_3$ is obtained just from the RG-running, 
independently of its initial value\footnote{Note, however, that the value
of $V_{33}$ in this case is not affected by the RG running and thus has to
be fixed by physics beyond $\Lambda$.}. This was noted in
ref.~\cite{cein3} for the (intermediate) case {\em (iii)} , but clearly
it also works for the completely degenerate case {\em (ii)}. Let us also
notice that the neutrinoless double $\beta$-decay constraint implies that 
case {\em (ii)} is only consistent if $m\simlt 0.2$ eV, see eq.~(\ref{B}).

\subsection{Neutrino masses}

The first consequence from eq.~(\ref{miRG}) is that  neutrino mass
differences get small modifications, unless the scenario is supersymmetric
and $\tan\beta \gg 1$, so that $\kappa_\tau \simgt {\cal O}(1)$.  This
means in particular that if the neutrino spectrum is hierarchical, \ie\
$m_1^2<m_2^2\ll m_3^2$, satisfying $\Delta m_{ij}^2\sim \max\{m_i^2,
m_j^2\}$, then radiative effects are not going to appreciably change the
spectrum and the mass splittings: $|\delta_{RGE} \Delta m^2_{ij}|\ll
|\Delta m^2_{ij}|$.

However, in a completely or partially degenerate neutrino scenario,
\ie\  $m_1^2\simeq m_2^2\simeq m_3^2$ or $m_1^2\simeq m_2^2\not\simeq
m_3^2$ respectively, the shifts induced by the RGEs may be of the order or
larger than the initial ones. They could also be larger than those
required to explain solar neutrino oscillations. As for the mixing
angles, the relevant cases in the study of the RGEs for the masses involve
some complete or partial degeneracy.  There is an important difference
however. We learnt from the previous  subsection that the mixing
angles can only appreciably change if some
$|\kappa_\tau\tilde\nabla_{ij}|\simgt 1$. For the mass splittings,
there can be cases where, even for much smaller
$|\kappa_\tau\tilde\nabla_{ij}|$ the RG effects are physically
relevant. Actually, they can be very important too if the signs of the
masses are opposite, $\tilde m_i\simeq -\tilde m_j$, in clear contrast 
with the mixing angles (notice that in this case 
$|\tilde\nabla_{ij}|\ll 1$).
In particular, starting from a high-energy scale, $\Lambda$, down to $M_Z$
the ``solar'' splitting $\Delta m^2_{21}$ can get a RG correction 
\bea 
\label{deltaRG}
\delta_{RGE} \Delta m^2_{21}\sim 4 m^2 \kappa_\tau \langle V_{32}^2 -
V_{31}^2\rangle \log{\frac{\Lambda}{M_Z}},
\eea 
where $\langle V_{32}^2 -V_{31}^2\rangle$ is to be understood as an average
value in the interval of running.  In degenerate scenarios
this is typically much larger than the splitting required for
the VO solution,
 unless $\langle V_{32}^2 - V_{31}^2\rangle \simeq 0$, \ie\ 
$ (s_1s_3 - c_1s_2c_3)^2 \simeq (-s_1c_3-c_1s_2s_3)^2$ or, equivalently
\cite{cein3,BRS},
\be
\tan 2\theta_3\simeq \frac{\cos^2\theta_1 \sin^2\theta_2
-\sin^2\theta_1}{\sin\theta_2\sin 2\theta_1}.
 \label{U31U32}
\ee 
This can be most naturally achieved with $s_2\simeq 0$, $\sin^2 2
\theta_3 \simeq 1$, which amounts to a scenario close to bimaximal mixing,
although it is not the only possibility. On the other hand, this way-out for
the VO solution requires that $V_{32}^2-V_{31}^2$ must be stable along the
RG running. This implies in turn that both $V_{32}$ and $V_{31}$ must be
stable. 
To see this notice that if $\tilde \nabla_{21}$ is dominant, then 
$V_{32}^2+V_{31}^2\simeq$ const., while if $\tilde \nabla_{32}$ is 
dominant, then $V_{31}^2=1-V_{32}^2-V_{33}^2\simeq$ const.
As discussed in the previous subsection, and it is apparent
from eq.~(\ref{URG}), $V$ will change little, unless some
$|\kappa_\tau\tilde \nabla_{ij}|\simgt 1$, which corresponds precisely to a
(at least partially)  degenerate case in which the $\tilde m_i$ and $\tilde
m_j$ masses have equal signs
($|\kappa_\tau\tilde \nabla_{ij}|$ could also be sizeable if the scenario is
supersymmetric with very large $\tan\beta$). Let us analyze all the 
possibilities.

If $\tilde m_1\simeq \tilde m_2$, \ie\ both have the same sign, then
$|\tilde \nabla_{21}|$ is the largest $|\tilde \nabla_{ij}|$. $|\tilde
\nabla_{21}|$ should not be large enough to lead the matrix $V$ to its
stable form,  as that would imply either $V_{31}\rightarrow 0$ or
$V_{32}\rightarrow 0$. But this means that $\Delta m_{21}^2$ cannot be too
small initially (or along the running). More precisely, it must be
$\Delta m_{21}^2\simgt 2|\epsilon_\tau |m^2$. In the case $m_1^2\simeq
m_2^2\gg m_3^2$ this is too large for the
VO solution since $m^2$ must be at least $\sim \Delta m_{atm}^2$. The
opposite case, with $m_1^2\simeq m_2^2\ll m_3^2$, can be viable but only
if the degenerate pair has very small masses, $m^2\simlt 10^{-6}\,
{\mathrm eV}^2$.

With that caveat in mind, if $m_1^2 \sim m_2^2$, the VO scenario requires
$\tilde m_1\simeq -\tilde m_2$, to avoid a disastrously large 
$|\tilde \nabla_{21}|$.
In that case, $|\tilde \nabla_{21}|\ll 1$ and
$\tilde \nabla_{32}$ (or equivalently $\tilde \nabla_{31}$ if it is
$\tilde m_1$ the one with the same sign as $\tilde m_3$)
will be dominant. Still we need from eq.~(\ref{deltaRG})
$\langle V_{32}^2-V_{31}^2\rangle\simeq 0$ with high accuracy, 
so one has to guarantee that the matrix $V$ 
does not change much, to avoid dangerous changes in $V_{32}^2-V_{31}^2$. 
In particular, to avoid running into the stable form of $V$ we require again
 $\Delta m^2_{32}\simgt  
2 m^2 \kappa_\tau \log{\frac{\Lambda}{M_Z}}$. In the present case, 
this condition is easily satisfied for realistic $\Delta m^2_{32}$, so
it is natural to get a quite stable matrix $V$, as required. 
But still, $V_{32}^2-V_{31}^2$ will (slightly) change according to
\bea
\frac{d (V_{32}^2-V_{31}^2)}{dt}\simeq 2\kappa_\tau V_{33}^2 V_{32}^2\tilde
\nabla_{32} .
\label{difRG3}
\eea
Then,  we can integrate eq.~(\ref{difRG3}) in leading log approximation
and substitute in eq.~(\ref{deltaRG}) to obtain
\bea
\delta_{RGE}\Delta m^2_{21}\sim -4 m^2 \kappa_\tau^2 V_{33}^2V_{32}^2 
\tilde \nabla_{32}\log^2{\frac{\Lambda}{M_Z}}\simeq -\frac{1}{2}m^2
\kappa_\tau^2
\tilde \nabla_{32}\log^2{\frac{\Lambda}{M_Z}}\ ,
\eea
(we have used $\langle V_{33}^2V_{32}^2\rangle\simeq 1/16$) which implies
\bea
\Delta m^2_{21}\equiv \Delta m^2_{sol} > 2\frac{m^4}{\Delta
m^2_{atm}}\kappa_\tau^2\log^2{\frac{\Lambda}{M_Z}} .
\label{m450} 
\eea
This condition can be satisfied, but is
barely compatible with a relevant cosmological role for neutrino masses
and the VO solution (for
the MSSM this conclusion is stronger as $\kappa_\tau$ is larger).
On the other hand,
condition (\ref{m450}) is easily fulfilled in a partially degenerate 
scenario (the so-called pseudo-Dirac or
intermediate case, $m_1^2\sim m_2^2\gg m_3^2$), where $m^2\sim \Delta
m^2_{atm}$.
Let us also remark
that the splitting induced by the RGEs is potentially so large as compared
to that required for the VO solution, that the constraint (\ref{U31U32})
must be satisfied with enormous accuracy [if some symmetry is invoked
\cite{BRS} to justify (\ref{U31U32}), it must be either exact or
minutely broken]. 

The previous paragraphs summarize and extend to the general case the
results from refs.\cite{cein1,cein2,cein3} concerning the mass splittings
induced by radiative corrections, especially in relation to the viability
of the VO solution.

\subsection{The two-flavour case}

It has been claimed \cite{ellis,ellos,babu} that, working in the two
flavour ($\nu_\mu, \nu_\tau$) approximation, the RG running could drive
the atmospheric angle $\sin^2 2\theta_1$  from a nearly minimal value at
high energy to nearly maximal at low energy. This is an interesting
possibility, since $\sin^2 2\theta_1$ is known to be  nearly maximal.
This subsection is devoted to this particular issue
on the light of the previous discussions in this section.

The analysis of refs.\cite{ellis,ellos,babu} was based on the RG equation
for $\sin^2 2\theta_1$ \cite{babu}  
\bea 
\frac{d }{dt}\sin^2 2\theta_1 = 2\kappa_\tau\ \sin^2 2\theta_1\cos^2
2\theta_1 
\frac{{\cal M}_{33}+{\cal M}_{22}}{{\cal M}_{33}-{\cal M}_{22}}.
\label{Babutheta}
\eea
The observation was that if the diagonal elements of the mass matrix are
close enough, $\sin^2 2\theta_1$ could undergo a large increment.

Now, starting from the general three flavour case, the two flavour 
approximation concerning RG running (not to be confused with the
two-flavour approximation for oscillations) will be exact if $s_2=s_3= 0$,
which is a stable condition. The matrix $V$ is then simply given by
\bea 
V\simeq \pmatrix{1 &  {}  &  {} \cr
{} &   c_1  &  s_1 \cr
{} & -s_1 &  c_1\cr}\ ,
\label{U2x2}
\eea 
There is a physically interesting  instance, namely the
SAMSW solution, which essentially corresponds  to this scenario.
Then, from (\ref{teta1}), the RGE for $\theta_1$ is
\bea 
\frac{d \theta_1}{dt} = \kappa_\tau\sin \theta_1 \cos \theta_1 \tilde
\nabla_{32}=
\kappa_\tau\sin \theta_1 \cos \theta_1
\frac{\tilde m_3+\tilde m_2}{\tilde m_3 -\tilde m_2}\ .
\label{thetanuestra}
\eea
The corresponding RG equation for $\sin^2 2\theta_1$ is then given by
\bea 
\frac{d }{dt}\sin^2 2\theta_1 = 2\kappa_\tau\sin^2 2\theta_1\cos
2\theta_1 
\frac{\tilde m_3+\tilde m_2}{\tilde m_3-\tilde m_2}\ .
\label{sin2tnuestra}
\eea
It is easy to check that eq.~(\ref{sin2tnuestra}) is indeed equivalent to
eq.~(\ref{Babutheta}), but it is more practical, as it is written in 
terms of physical parameters. 
In any case, eq.~(\ref{thetanuestra}) is 
the more useful equation for the analysis\footnote{
Incidentally, $\sin^2 2\theta_1=1$, \ie\ maximal, is not 
a stable RG point, as has been argued in the literature  
\cite{babu}. From the general equations 
(\ref{teta1}-\ref{teta3}) we can see that this is also the 
generic case in the 3-flavour scenario, although one can easily 
determine particular conditions for stability. {\it E.g.} if 
$\tilde \nabla_{32}$ is the dominant term and $c_3=0$, 
then $\theta_1$ is stable for any value.}.
Clearly, we can only expect a large modification
in the angle if $|\kappa_\tau\tilde \nabla_{32}|
\simgt 1$. This requires $\tilde m_3\sim \tilde m_2$, unless the scenario
is supersymmetric with very large $\tan\beta$ ($\simgt 100$). 
So, we must be in
case {\em (ii)} (note that case {\em (iii)} has $c=3$ by definition).
However, our conclusions were that case {\em (ii)} can only be
acceptable if labels $a$, $b$ correspond to 1, 2, and thus $c$ to 3, 
which is inconsistent with $\tilde m_3\sim \tilde m_2$.
The  solution to this apparent contradiction is that 
the viability of this $\tilde m_3\sim \tilde m_2$ scenario
requires the running to stop before reaching the stability form of $V$,
while in the analysis of subsection 3.1 we assumed that the 
stable form was reached. Let us see this in closer detail.

If we start for example
with $\theta_1\simeq 0$ (but not exactly $\theta_1= 0$, which is a
stable value) and negative $\kappa_\tau\tilde \nabla_{32}$, the 
RGE (\ref{thetanuestra}) will drive
$\theta_1\rightarrow\pi/2$. Schematically,
\bea 
\label{schem}
\pmatrix{1 &  {}  &   {} \cr
{} &   1  &  0 \cr
{} & 0 &  1\cr}\ \rightarrow 
\pmatrix{1 &  {}  &   {} \cr
{} &   \frac{1}{\sqrt{2}}  &  \frac{1}{\sqrt{2}} \cr
{} & -\frac{1}{\sqrt{2}} & \frac{1}{\sqrt{2}} \cr}\ \rightarrow 
\pmatrix{1 &  {}  &   {} \cr
{} &   0  &  1 \cr
{} & 1 &  0\cr}
\eea 
(changing the sign of $\kappa_\tau\tilde \nabla_{32}$ would reverse 
the direction of the arrows).
Notice that this is in agreement with our conclusions for case 
{\em (ii)}: if $|\kappa_\tau\tilde \nabla_{32}|\simgt 1$,
$V_{33}\rightarrow 0$, which is inconsistent with experiment.
However, there
is an intermediate scale at which the mixing must be maximal, $\sin^2
2\theta_1=1$ (of course this does not occur if we start already near the
stable $\theta_1=\pi/2$ value).  This critical scale may be at low
energy, and this is the possibility exploited in 
ref.\cite{ellis,ellos,babu}. This
of course requires some tuning of the initial  parameters, and we will
be more explicit about this shortly. 

Comparison of eq.~(\ref{thetanuestra}) with the general equation for
$\theta_1$, eq.~(\ref{teta1}), shows that there could be other ways in
which the two-flavour approximation for the running is correct. Namely, 
 eq.~(\ref{teta1}) reduces to (\ref{thetanuestra})  
if one of the following possibilities occur
\bea
&(a)&\hspace{1cm} s_3=0, \nonumber\\
&(b)&\hspace{1cm} \tilde \nabla_{31}\simeq \tilde \nabla_{32}.\nonumber
\eea
These cases were not apparent at all from eq.~(\ref{Babutheta})
used in previous analyses. Now, scenario {\em (a)} is only acceptable
if the $s_3=0$ condition is stable along the running, while $s_1$ is 
substantially changing. It can be
checked from (\ref{teta1}--\ref{teta3}) that this occurs only if
$s_2=0$. Therefore, scenario {\em (a)} can only work in the 
case previously commented, \ie\ eq.~(\ref{U2x2}).  On the other
hand, scenario {\em (b)} can work in many cases. 

\begin{figure}
\centerline{\vbox{
\psfig{figure=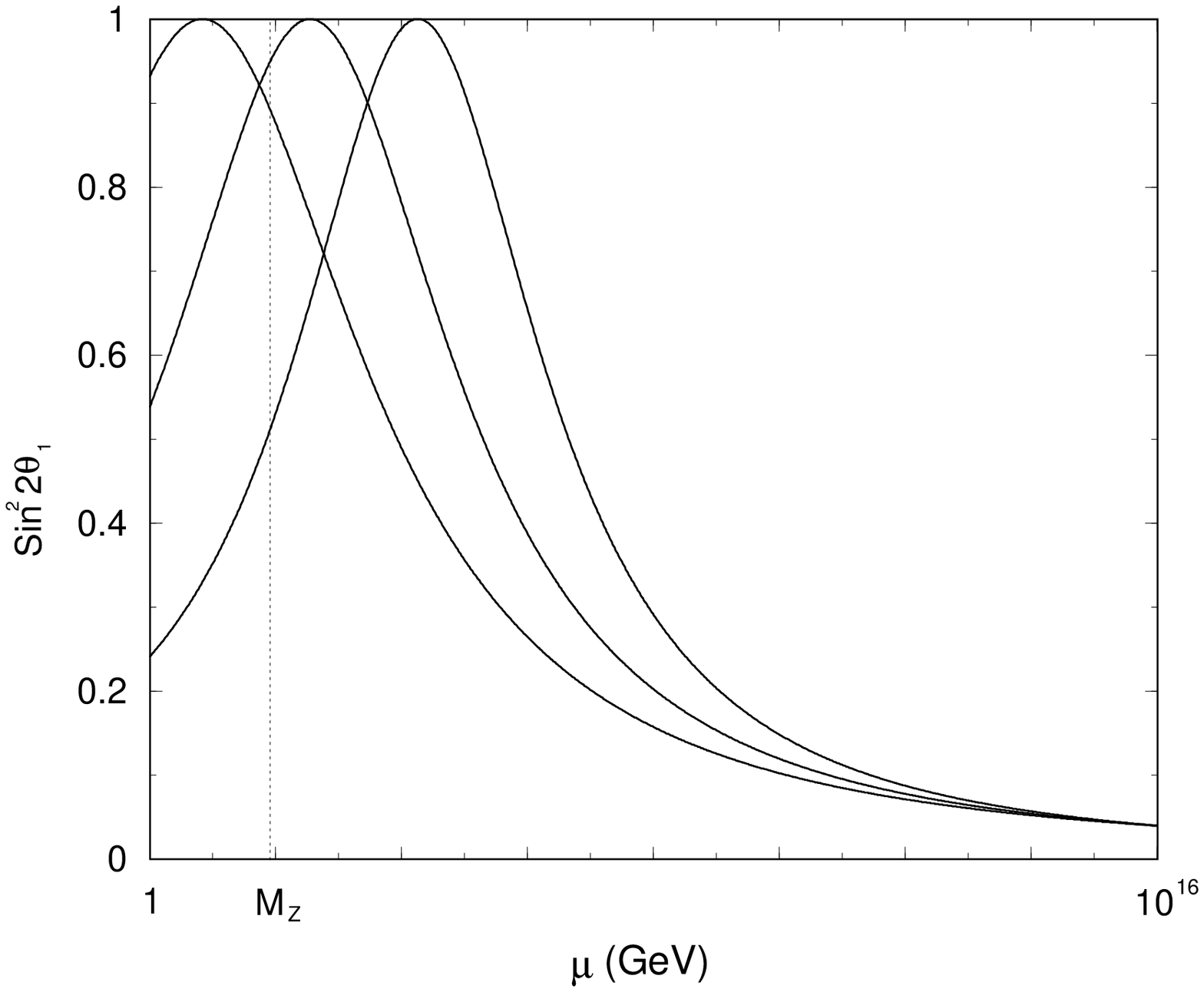,height=10cm,width=10cm,bbllx=3.cm,%
bblly=1.cm,bburx=18.cm,bbury=16.cm}
\psfig{figure=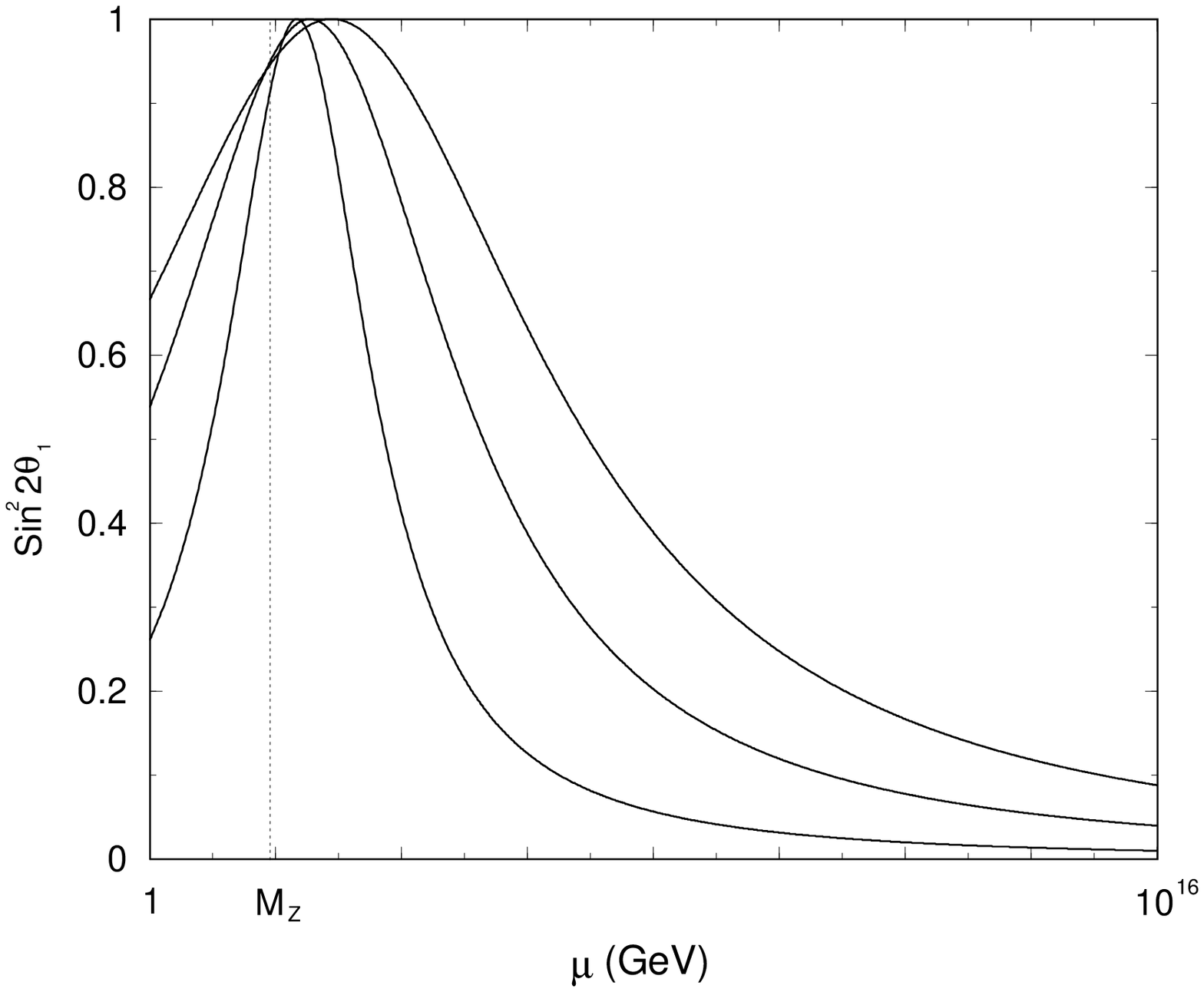,height=10cm,width=10cm,bbllx=3.cm,%
bblly=1.cm,bburx=18.cm,bbury=16.cm}}}
\caption
{\footnotesize 
Running of $\sin^22\theta_1$ in the two-flavour approximation.
In the upper plot, we fix the initial value of the mixing angle but vary
the initial mass splitting $\Delta m= 0.00393$ eV in $\pm 0.0005$ eV.
In the lower plot, we vary instead the initial value of the angle.
We have chosen $m=1$ eV and $\kappa_\tau=6.2\times 10^{-5}$ to
get maximal mixing near $M_Z$. }
\end{figure}

Let us now estimate the amount of fine-tuning that these two-flavour 
scenarios require for the the atmospheric angle to be driven to maximal
values at $M_Z$ thanks to the RG running.  
In particular, for case {\em(a)}, we notice that the
product  $V_{32}V_{33}\propto \sin 2\theta_1$ must grow 
($V_{32}V_{33}\rightarrow
V_{32}V_{33}/F$) in a factor 
$1/F$,  with $F\ll 1$.  From
the discussion in subsection 3.1, 
we know that this requires a suitable cancellation in eq.~(\ref{deltaRGE3})
between the initial mass splitting $\Delta m^2$ and that induced by the
running $2\epsilon_\tau m^2$.
Assuming for example a supersymmetric scenario (thus $\kappa_\tau<0$)  
with $\tilde \nabla_{ij}>0$, we find
\bea  
\left|
\frac{\Delta m^2 - 2\epsilon_\tau m^2}{\Delta m^2}
\right|<F\ll 1,
\label{fine2x2}
\eea 
(we have made the estimate $\langle V_{33}^2-V_{32}^2 \rangle\sim 1/2$,
which works well for small values of the initial mixing, as we have checked 
numerically).
This shows that the fine-tuning between the two terms in the numerator
of (\ref{fine2x2}), which have completely different origins,
is of one part in $1/F$.  Therefore, the more
important the increase in  $\sin^2 2\theta_1$, the greater the fine-tuning. 
Figure 1 illustrates well the behaviour of the running $\sin^2 2\theta_1$
and the fine-tuning involved. We show $\sin^2 2\theta_1$
as a function of the energy scale $\mu$, from $\Lambda=10^{16}$ GeV down to
$M_Z$ (we let the angle run below $M_Z$, where it should really stop, just
for illustrative purposes) according to eq.~(\ref{thetanuestra}). The
parameters
have been chosen to reach maximal mixing near $M_Z$ starting from a small
mixing angle at $\Lambda$. It is shown how, after reaching that maximal
value, $\sin^22\theta_1$ goes down again, seeking the stable 0 value [in
accordance with the expected behaviour, see (\ref{schem})]. In the upper
plot we demonstrate the dependence of the effect on the initial mass
splitting: when it is varied slightly (leaving the initial angle fixed),
the scale at which maximal mixing is achieved moves quite rapidly: an
initial mass splitting slightly off and the mixing at $M_Z$ drops below the
experimental lower bound. 
%
%
The lower plot shows instead the dependence with
the initial condition for the mixing. When that initial value increases
(decreases), the fine-tuning required is smaller, and this is reflected in
a wider (thinner) half-width of the curve.

Still within case {\em (a)}, 
using again the conservation of $V_{33}V_{32}/\tilde\nabla_{32}$,
we can easily evaluate the relation between the initial and final mass
splittings as
\be
\Delta m^2 (M_Z) \equiv \Delta m^2_{atm} \simeq F\  \Delta m^2 (\Lambda). 
\ee
Using eq.~(\ref{fine2x2}) we can obtain what is the value of $\kappa_\tau$
required as
\be 
|\kappa_\tau| \sim \frac{1\pm F}{2F}\frac{\Delta
m^2_{atm}}{m^2 \log\frac{\Lambda}{M_Z}} .
\label{kappa}
\ee
For example, if $F\sim 1/5$, which implies an important increase 
in $\sin^22\theta_1$ with a moderate fine-tuning, and using 
 $10^{-3}\ {\rm eV}^2 \leq m^2 \leq 4 \ {\rm eV}^2$, $\Delta
m^2_{atm}=10^{-3}$ eV$^2$ and $10^{5}\ {\rm GeV}
\leq \Lambda \leq 10^{16} \ {\rm GeV}$, we obtain $1.5\times 10^{-5} \simlt
|\kappa_\tau | \simlt 0.28$, which means that the scenario must be
supersymmetric with $\tan\beta>5$. 

On the other hand, since $V_{32}$ and $V_{33}$ are not stable, 
we can evaluate from eq.~(\ref{deltaRG}) the increment in the
``solar'' neutrino mass splitting $\Delta m^2_{21}$. Using
now $\langle V_{32}^2 - V_{31}^2\rangle\sim F/2$ and (\ref{kappa}),
we get
\bea \label{largerg}
\delta_{RGE}\Delta m^2_{21}\sim 4 \Delta m^2_{atm}
\eea
which is too large. This is not surprising, as $m_1^2$ is stable but 
$m_2^2$ experiments a change of order $\Delta m^2_{atm}$. Of course,
one could still get $\Delta m^2_{21}$ of the right size tuning the value
of $\tilde m_1$ so as to compensate the large RG correction (\ref{largerg})
but that would be rather unnatural. So, we conclude that the scenario {\em
(a)}, or equivalently eq.~(\ref{U2x2}), cannot implement a substantial
increase of $\sin^22\theta_1$ in practice.

Let us consider now scenario {\em (b)}. In principle, it can be
acomplished in many cases. Namely, whenever $m_1^2\simlt m_2^2\ll
m_3^2$ (hierarchical spectrum); $m_1^2\simeq m_2^2\gg m_3^2$
(intermediate or pseudo-Dirac spectrum); or $m_1^2\simeq m_2^2\simeq m_3^2$
(degenerate spectrum) with $\tilde m_1/\tilde m_2>0$ and $\Delta
m^2_{21}\ll\Delta m^2_{32}$. In the first two cases, $\tilde
\nabla_{32}\simeq -1$. Therefore, if one demands a significant
increase of $\sin^22\theta_1$ in eq.~(\ref{thetanuestra}), then one
needs  $|\kappa_\tau|\log\frac{\Lambda}{M_Z}\simgt 1$,  which signals the
breakdown of perturbation theory.  With regard to the degenerate
case, since $\tilde m_1$ and $\tilde m_2$ have equal signs, the
neutrinoless double $\beta$-decay constraint implies  that $m\simlt
0.2$ eV, see eq.~(\ref{B}). This means that $|\tilde \nabla_{32}|$
cannot be very large. Taking $\Delta m^2_{atm}>10^{-3}\ {\rm eV}^2$,
we get $|\tilde \nabla_{32}|=4 m^2/\Delta m^2_{atm}\simlt
160$. Therefore, an appreciable variation of $\sin^22\theta_1$
requires $|\kappa_\tau|\log\frac{\Lambda}{M_Z}\sim \Delta m^2_{atm}/m^2
\simgt 1/40$, which
means $\tan\beta>50$, quite a large value. Anyway, in this
degenerate case, $|\tilde \nabla_{21}|\gg|\tilde
\nabla_{32}|$. Consequently, if $|\tilde \nabla_{32}|$ is large enough
to modify the $V_{3i}$ entries (which is our assumption), $\tilde
\nabla_{21}$ will lead to $V_{31}\rightarrow 0$ (or equivalently
$V_{32}\rightarrow 0$) rapidly, which requires an SAMSW scenario
[see discussion after eq.~(\ref{Ustable2})]. But this is disastrous:
from eq.~(\ref{deltaRG}), noting that $\langle
V_{32}^2-V_{31}^2\rangle$ is necessarily sizeable,
$|\delta_{RGE} \Delta m^2_{21}|= {\cal O}(\Delta m^2_{atm})$,
in disagreement with observations (again barring an unnatural
cancellation between the initial and the RG-induced mass splitting).

To summarize, the two-flavour approximation for the running of 
the $\theta_1$ angle can
be a good one if the previous $(a)$ or $(b)$ conditions are fulfilled.
Then the size of $\kappa_\tau \tilde \nabla_{32}$ will determine whether
$\sin^22\theta_1$ can get a substantial increase or not. If it can, then
$\Delta m_{21}^2$ will get too large along the running.

The possibility of obtaining an enhancement of
$\sin^22\theta_1$ by the RG running seems so attractive that one may
wonder if it could be realized in a 3-flavour scenario. Unfortunately,
things do not seem to improve much.
From the previous discussion, it is clear that the only possibility
to consider here is $m_1^2\simeq m_2^2\simeq m_3^2$ (degenerate spectrum)
with $\tilde m_1/\tilde m_2<0$ and $\Delta m^2_{21}\ll\Delta
m^2_{32}$. Otherwise, the scenario is equivalent to the two-flavour
case, which we know does not work, or it has too small
$\kappa_\tau \tilde \nabla_{ij}$ to produce appreciable modifications
in the angles. In the scenario selected only 
$\kappa_\tau \tilde \nabla_{32}$ can be relevant.
The effect of the running on the $\Delta m_{21}^2$ splitting is still given by
eq.~(\ref{deltaRG}), which means that we will get 
$|\delta_{RGE} \Delta m^2_{21}|= {\cal O}(\Delta m^2_{atm})$
(unacceptable in any scenario) unless $V_{32}^2 -V_{31}^2\sim 0$.
This condition can be certainly arranged at low energy (\ie\ at the final
point of the running) using $s_2=0$, $\sin^22\theta_3=1$ 
(which implies a nearly bimaximal scenario). However, 
if the running is going to modify appreciably the matrix $V$ 
(which is mandatory to get an important enhancement of
$\sin^22\theta_1$) this condition will not be stable.
Notice that the dominance of $\kappa_\tau \tilde \nabla_{32}$
imply that $V_{31}\simeq $ const., while $V_{32}, V_{33}$ 
should vary substantially. In consequence the final 
$\Delta m^2_{21}$ will be again too large.

Schematically, the modification of the matrix $U$ (starting with 
low $\sin^22\theta_1$) will be 
\bea 
\label{schem2}
\pmatrix{\frac{1}{\sqrt{2}} & -\frac{1}{\sqrt{6}}   &  \frac{1}{\sqrt{3}}
\cr
-\frac{1}{2} &   \frac{1}{2\sqrt{3}}  &  \frac{2}{\sqrt{6}} \cr
\frac{1}{2} & \frac{\sqrt{3}}{2}  &  0\cr}\ \rightarrow 
\pmatrix{\frac{1}{\sqrt{2}} & \frac{1}{\sqrt{2}}  &   0 \cr
-\frac{1}{2} &  \frac{1}{2}  &  \frac{1}{\sqrt{2}} \cr
\frac{1}{2} & -\frac{1}{2} & \frac{1}{\sqrt{2}} \cr}\ \rightarrow 
\pmatrix{\frac{1}{\sqrt{2}} &  \frac{1}{\sqrt{3}} & -\frac{1}{\sqrt{6}} \cr
-\frac{1}{2}   &  \frac{2}{\sqrt{6}} &   \frac{1}{2\sqrt{3}} \cr
\frac{1}{2}  &  0 & \frac{\sqrt{3}}{2} \cr}.
\eea 
(This example has $\cos\theta_1\sim0$, but the conclusions are the same for
examples with $\sin\theta_1\sim0$). In the
initial part of the running $\sin^22\theta_1$ experiments
a large enhancement, so the parameters 
($\kappa_\tau \tilde \nabla_{32}$ and $\Lambda$) should be tuned 
to have the running finishing at the intermediate point. But in any case,
it is apparent that $V_{32}^2 -V_{31}^2\neq 0$ along the running, 
thus yielding too large $\Delta m^2_{21}$.

\section{Implications (general complex case)}

Non-zero CP-violating phases can have a non-trivial effect on the RG
evolution of neutrino mixing angles and masses. Such effects have been
considered previously in a 2 generation case \cite{haba}. With the
formalism presented in section 2 we can undertake easily the more realistic
analysis for 3 generations.

As in the previous section, we will assume that below the high-energy scale
 $\Lambda$, the effective theory is just the SM or
the MSSM, plus the effective  Majorana mass matrix for the neutrinos.
As shown in sect.~2, the RGEs for the neutrino masses and the matrix $U$
are given by eqs.~(\ref{RGmass}, \ref{RGU}), with the appropriate 
substitutions. Explicitly,
\be
\label{RGmass2}
\frac{d m_i}{dt}= - 2 m_i |U_{3i}|^2 - m_i \kappa_U,
\ee
\be
\label{RGU2}
\frac{d U}{dt}= U T,
\ee
where $\kappa_U$ gives a universal effect 
[see eqs.~(\ref{KSM}, \ref{KMSSM})] and 
\bea 
T_{ii}&=&0\nonumber\vspace{0.1cm}\\
T_{ij}&=&\kappa_\tau\left[\nabla_{ij}Re
(U_{3i}^*U_{3j})+i\nabla_{ij}^{-1}Im (U_{3i}^*U_{3j})\right],\,\, (i\neq
j).
\eea 
The explicit RGEs for the mixing angles, the
CP-violating
phases and the unphysical phases are given by
eqs.~(\ref{RGt1}--\ref{alfatau}),
using everywhere the substitutions
of eqs.~(\ref{PSMdef}, \ref{TSMdef}).

As in the real case, large effects in the
radiative corrections are expected when the neutrino spectrum has
(quasi)-degenerate eigenvalues. When the degeneracy is sufficiently good
(see previous section), $U$ changes rapidly from
its unperturbed $t=0$ form to a stable form that has
non-singular $dU/dt$. To be
more precise, if $m_i\simeq m_j$, then $|\nabla_{ij}|\gg 1$, and $U$ is
rapidly driven to a form that ensures $T_{ij}\simeq 0$. 
Since near $t=0$, one has $T_{ij}\sim
\nabla_{ij} Re (U_{3i}^*U_{3j})$, the stable form, say $U'$, 
must satisfy
\bea
\label{stable}
Re ({U'}_{3i}^*U'_{3j})=0
\eea
for any pair $i,j$ with $m_i\simeq m_j$. Since the unphysical phase 
$\alpha_\tau$, cancels out in the previous equation, we can also write
\be
\label{RUIUJ}
Re ({W'}_{3i}^*{W'}_{3j})=0,
\ee
where the matrix $W$ was defined in eq.~(\ref{W}).

If all masses are quite different, then no dramatic RG effects appear and
the
discussion is similar to the corresponding one in the real case (see
previous section). In the following subsections we then consider the two
interesting cases of two-fold or three-fold degeneracy. The following
general analysis contains as a particular case the real scenario considered
in detail
in the previous section. Note that now, two real scenarios with degenerate
masses of either equal or opposite signs are treated together and correspond
to choices of $\phi$ (or $\phi'$) equal to 0 or $\pi$, respectively. Note
that, although the case with opposite signs has now a very
large $|\nabla_{ij}|$, having $\phi-\phi'=\pi$ gives a $T_{ij}$ under
control, and the case is, of course, stable.

As in the real case, for the discussion of cases of physical relevance, we
will assume that  the solar neutrino problem is solved by one of the
standard solutions, with $|\Delta m^2_{21}|\ll  |\Delta m^2_{32}| \sim
|\Delta m^2_{31}|$.

\subsection{Three-fold degeneracy}

In the case $m_1\simeq m_2\simeq m_3\simeq m$, with  
$|\kappa_\tau\nabla_{ij}|\gg 1$
for all $i,j$, the matrix $U$ (or equivalently $W$)
quickly reaches a stable value $U'$ ($W'$) with
\be
\label{threefold}
Re ({W'}_{31}^*{W'}_{32})=Re ({W'}_{32}^*{W'}_{33})=Re
({W'}_{31}^*{W'}_{33})=0.
\ee
Using the fact that ${W'}_{33}$ is real in the parametrization
(\ref{Vdef}),
the last two equalities in (\ref{threefold}) imply
\be
W'_{33}=0\;\;\;\; {\mathrm or}\;\;\;\; Re(W'_{31})=Re(W'_{32})=0.
\label{Wp33}
\ee
If $W'_{33}\neq 0$, eq.~(\ref{Wp33}) and 
the first equality in
(\ref{threefold}) imply
\be
W'_{31}=0,\;\;\;\; {\mathrm or}\;\;\;\; W'_{32}=0.
\ee
In other words, there must be some $W'_{3i}=0$ (the ambiguity in
the labelling of the three original eigenvalues is reflected in the
ambiguity in the $i$-label of $W'_{3i}=0$).
Then, by unitarity, the values of $|W'_{3j}|^2=|U'_{3j}|^2$ are
$\{0,x,1-x\}$, with $0\leq x\leq 1$ 
(the value of $x$ is determined by the initial form of the 
matrix $W$). So, from the expressions for $d m_i/dt$
we obtain the following low-energy shifts in the mass eigenvalues:
\be
\Delta m_i\simeq m \epsilon_\tau\{0,x,1-x\},
\ee
where $m$ is essentially the initial value of the three masses at 
the scale $\Lambda$. 

Therefore, for a given value of $x$ we know how the degeneracy 
in the mass spectrum
is lifted and then the labelling of
the three eigenvalues is determined according to our convention (thus $U'$
is fixed). It is then easy
to show that no value of $x$ gives a stable $U'$ in agreement with
the available experimental information on the neutrino mixing angles.
For $x<0.25$ (or $>0.75$) one has $U'_{31}\rightarrow 0$ and
$|U_{33}|^2>0.75$, in conflict with the limit $c_1^2c_2^2<0.71$ which
follows from the experimental bound (\ref{atmexp}). For $0.25<x<0.75$,
$U'_{33}\rightarrow 0$ instead, and this is also incompatible with
experimental limits: it implies either $c_2^2\rightarrow 0$ [in conflict
with (\ref{exp2})], or $c_1^2\rightarrow 0$ [in conflict with
(\ref{atmexp})].

In summary, the case of three-fold degeneracy of initial neutrino
masses  leads to a stable $U'$ which does not accomodate the values of
$\theta_i$ and $\Delta m_{ij}^2$ suggested by experiment. Note that this
analysis contains
and expands the subcases {\em (i)--(ii)} for real $U$ analyzed in the
previous section.  This negative result has two obvious
way-outs. If the spectrum is almost degenerate, but all
$|\kappa_\tau\nabla_{ij}|\ll 1$, then the matrix $U$ is very slightly
renormalized, so it never reaches the stable (disastrous) form.  In
this case the RG corrections to mixing angles and phases are basically
irrelevant.  This extreme situation is difficult to implement in the
VO scenario,  since  $|\kappa_\tau \nabla_{21}|$ is necessarily sizeable.
The second way-out is that $|\kappa_\tau\nabla_{32}|\ll 1$ (even though
$|\nabla_{32}|\gg 1$), while $|\kappa_\tau\nabla_{21}|$ can be
significant. This will occur if the  $\Delta m^2_{32}$ splitting is
arranged to sensible (``atmospheric'')  values from the
beginning. Then, the analysis for the mixing angles and CP phases is
identical to a case with two-fold degeneracy, which is analyzed in
the next subsection.

\subsection{Two-fold degeneracy}

We assume here that the initial $\Delta m^2_{32}, \Delta m^2_{31}$
splittings are large enough to keep $m_3$ as the most split eigenvalue
after the RG effects, while $m_1\simeq m_2$.
In this case,
only $|\kappa_\tau\nabla_{21}|\simgt 1$. This drives the original matrix 
$U$ to a stable form $U'$ given by (see Appendix)
\be
\label{Gamma}
U'=UR_{12}(\Gamma),
\ee
where $R_{12}(\Gamma)$ is a rotation in the 1-2 plane by an angle
$\Gamma$, to be computed below.
We see that the third column of $U$ is not changed appreciably, in
accordance with the fact that $d U_{i3}/dt$ does not depend on
$\nabla_{21}$. The rotation angle $\Gamma$ is determined by the condition
(\ref{stable})
\be
Re ({U'}_{31}^*U'_{32})=0,
\ee 
which gives 
\be
\tan2\Gamma=\frac{2Re (U_{31}^*U_{32})}{|U_{32}|^2-|U_{31}|^2}.
\ee
This allows to determine $U'$ completely  and thus the physical parameters
as a function of the initial conditions. Let us examine this issue in 
closer detail.

With regard to the mixing angles, from the RGEs (\ref{RGt1}--\ref{RGt3}) one
immediately sees that $\theta_1$ and $\theta_2$ are not renormalized
strongly ($d\theta_{1,2}/dt$ do not depend on $T_{12}$) while $\theta_3$ is
driven to a stable value, $\theta_3^{(f)}$. More precisely, 
\bea
\label{theta3n}
\theta_1^{(f)}&\simeq& \theta_1\nonumber\\
\theta_2^{(f)}&\simeq& \theta_2\nonumber\\
\sin^22\theta_3^{(f)}&=&p^2+q^2,
\eea
where
\be
p=\sin2\Gamma\sin[(\phi-\phi')/2],
\ee
and 
\be
q=\sin2\Gamma\cos2\theta_3\cos[(\phi-\phi')/2]+\sin2\theta_3\cos2\Gamma.
\ee
The CP-violating phases are also driven to particular values $\delta^{(f)},
\phi^{(f)}, \phi'^{(f)}$ which are
complicated functions of the initial values of the parameters in $U$:
\be
\label{deltan}
\tan\delta^{(f)}=\frac{q\tan\delta-p}{q+p\tan\delta}\ ,
\ee
\be
\label{phin}
\tan[(\phi^{(f)}-\phi'^{(f)})/2]=
\frac{\sin2\theta_3\sin[(\phi-\phi')/2]}{\cos2\Gamma\sin2\theta_3
\cos[(\phi-\phi')/2]+\sin2\Gamma\cos2\theta_3}\ ,
\ee
and 
\be
\label{con1}
\phi^{(f)}+\phi'^{(f)}-2\delta^{(f)}=\phi+\phi'-2\delta.
\ee
The last equation follows from $d(\phi+\phi'-2\delta)/dt\simeq 0$,
which can be checked directly from (\ref{RGCP1})-(\ref{RGCP3}). 

As explained in section 2, see
(\ref{CONS}-\ref{CONS3}),
other quantities of interest which are conserved in this case are 
$Im({U_{k1}^*U_{k2}})$. For $k=1$, we obtain the conserved quantity
\be
\label{con2}
\sin2\theta_3\sin[(\phi-\phi')/2],
\ee
and, for $k=2,3$, 
\be
\label{con3}
\sin[(\phi-\phi')/2]\cos\delta\cos2\theta_3
-\sin\delta\cos[(\phi-\phi')/2].
\ee
These expressions can be sometimes conveniently used instead of the
explicit formulas (\ref{theta3n})-(\ref{phin}).

For physical applications it is interesting to use for $\theta_1$ 
and $\theta_2$ the values suggested by
experiment\footnote{Due to the stability of $\theta_1$ and $\theta_2$, 
this initial
condition at $\Lambda$ is approximately maintained by RG evolution down 
to $M_Z$. The origin of such initial
condition could be 
some flavour symmetry or perhaps  a fixed point
in the RG running above $\Lambda$.}, $\sin^22\theta_1\sim 1$ and
$\theta_2\sim 0$, which permits to obtain more explicit
expressions for $\theta_3$ and the CP-violating phases
 at low energy. In
doing this, it is important to keep corrections due to a small (but
non-zero) value of $\theta_2$, as they can become dominant for some
choices of the parameters.
For the angle $\theta_3^{(f)}$, we obtain the remarkably
simple expression
\be
\label{theta3exp}
\sin^22\theta_3^{(f)}=\sin^22\theta_3\sin^2[(\phi-\phi')/2]+{\cal O}(s_2^2),
\ee
which gives the final $\theta_3^{(f)}$ as a function of the initial
$\theta_3$ and $\phi-\phi'$. We see that, for the SAMSW solution of the
solar neutrino problem, we must have either
$\sin^22\theta_3\sim 0$ or $\sin^2[(\phi-\phi')/2]\sim 0$ at the scale
$\Lambda$. On the other hand, if the solar mixing is also maximal, 
$\sin^22\theta_3^{(f)}\sim 1$, then both $\sin^22\theta_3$ and
$\sin^2[(\phi-\phi')/2]$ must be of order 1 originally. 

A word of caution should be said about the higher order term in
(\ref{theta3exp}). The neglected term, calculated exactly, is
[we introduce the short-hand notation $\varphi\equiv(\phi-\phi')/2$ and
 $\sigma\equiv \sin 2\theta_1=\pm 1$]:
\be
\label{neglect}
4s_2^2\frac{N_1^2}{D_1^2+D_2^2},
\ee
with
\bea
N_1&=&c_\varphi c_\delta+s_\varphi s_\delta c_{2\theta_3},
\nonumber\vspace{0.3cm}\\
D_1&=&-c_\varphi c_2^2 s_{2\theta_3}+2\sigma s_2(c_\varphi c_\delta
c_{2\theta_3}+s_\varphi s_\delta),
\nonumber\vspace{0.3cm}\\
D_2&=&c_2^2c_{2\theta_3}+2\sigma
s_2c_\delta s_{2\theta_3} \ ,
\eea
where all the quantities are understood at the initial (high energy) scale.
For generic values of the parameters, the term (\ref{neglect}) will be
negligible as it is formally ${\cal O}(s_2^2)$. However, for some
particular choices of the parameters the denominator in (\ref{neglect})
can be also ${\cal O}(s_2^2)$ resulting in a non-suppressed correction in
(\ref{theta3exp}). This occurs, for example, if $c_{2\theta_3}\simeq
0$ and $c_\varphi\simeq 0$ simultaneously.

Expressions (\ref{theta3exp}) and (\ref{neglect}) for $\sin^22\theta_3^{(f)}$,
in conjunction with the conserved quantities (\ref{con1}--\ref{con3}), are
enough to determine exactly the rest of parameters at low-energy.
For example, we obtain
\be
\label{varphin}
\tan^2[(\phi^{(f)}-\phi'^{(f)})/2]=
\frac{\sin^22\theta_3\sin^2\varphi (D_1^2+D_2^2)}{4 \sin^2\theta_2 N_1^2},
\ee
Note the $1/s_2^2$ dependence that would generically
drive $\phi^{(f)}-\phi'^{(f)}\rightarrow \pm \pi$. However, this is not
necessarily the case because the smallness of $s_2$ can be compensated by a
small numerator in (\ref{varphin}), as commented above.

If $\phi=\phi'$ initially, it is easy to see that the only
physical parameter that undergoes a sizable change is $\theta_3$, 
with $\theta_3^{(f)}=\theta_3+\Gamma$. This 
follows directly from $U'=UR_{12}(\Gamma)$ because now $R_{12}(\Gamma)$
commutes with ${\mathrm diag}(e^{i\phi},e^{i\phi},1)$.
Moreover, eq.~(\ref{theta3exp})
gives $\theta_3^{(f)}\rightarrow {\cal O}(s_2)\sim 0$ irrespective of the
initial value of
$\theta_3$.
This is just right for the SAMSW solution
of the solar neutrino problem and reproduces the result discussed
in the previous section for the real case. We see however that this appealing 
possibility dissappears for generic values of $\phi$ and $\phi'$.

If $\phi-\phi'=\pm\pi$ instead, one gets
$\tan2\Gamma\sim s_2\sin\delta$. For $\theta_2=0$ or $\delta=0$
this implies a stable matrix $U$
from the beginning, see eq.~(\ref{Gamma}). This is in correspondence with 
the
real
case $\phi=0, \phi'=\pi$ discussed in the
previous section. For non-zero values of $\delta$ and/or $\theta_2$, 
however,
the matrix $U$ is no longer stable, although typically $\Gamma$, and thus
the corrections to $U$, will be
small. We conclude that, 
the presence of a non-zero phase $\delta=0$ tends to destabilize 
this scenario, but the effect is small for small $\theta_2$.

Another interesting result that we can extract from eq. (\ref{deltan}) is
the possiblility of generating a non-zero phase $\delta^{(f)}$ even if
$\delta=0$ originally, provided the phases $\phi, \phi'$ are non zero and
satisfy $\sin[2(\phi-\phi')]\neq 0$. The
resulting $\delta^{(f)}$, when $\delta=0$, is given by
\be
\tan\delta^{(f)}=-\frac{p}{q},
\ee 
or, explicitly,
\be
\tan\delta^{(f)}=\frac{-s_\varphi c_\varphi (c_2^2\sin2\theta_3-2\sigma
s_2\cos2\theta_3)}{
c_2^2s_\varphi^2\sin2\theta_3\cos2\theta_3+2\sigma s_2
(1-s_\varphi^2\cos^22\theta_3)}
\ee
This shows that one can generate radiatively even a maximal value for the
CP-violating phase.

\subsection{Neutrino masses}
\begin{figure}
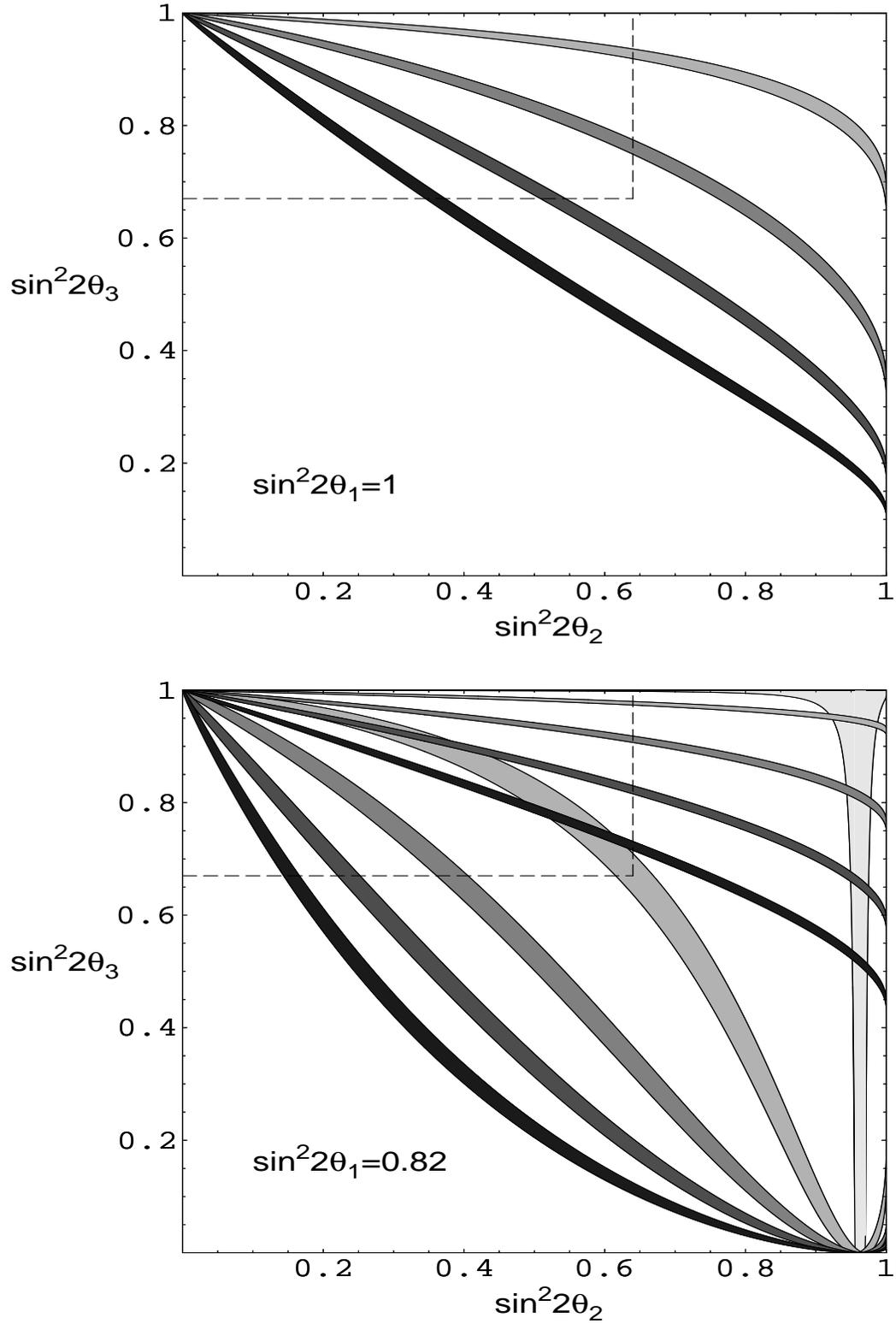

\centerline{\vbox{
\psfig{figure=cein4.2a.ps,height=10cm,width=10cm,bbllx=7.cm,%
bblly=5.cm,bburx=15.cm,bbury=15.cm}
\psfig{figure=cein4.2b.ps,height=10cm,width=10cm,bbllx=7.cm,%
bblly=5.5cm,bburx=15.cm,bbury=15.5cm}}}
\caption
{\footnotesize Allowed regions (in which $\Delta m_{21}^2\leq 1.1\times
10^{-10}\, {\rm eV}^2$) in the plane ($\sin^22\theta_2,\sin^22\theta_3$)
for the SM case for $\sin^22\theta_1=1$ (upper plot) and 0.82 (lower)
and a cut-off scale $\Lambda=10^{12}\, {\rm GeV}$.
The different strips correspond to
different values of the phase $\delta$, from 0 (darker color) to $\pi/2$
(lighter).}
\end{figure}

In the complex case, the running of the neutrino mass eigenvalues is
governed by
\be
\label{dmcomplex}
\frac{d m_i}{dt}=-2\kappa_\tau m_i|U_{3i}|^2-\kappa_U m_i,
\ee
and many of the generic results obtained in the real case go over to the
complex one. Some of the significant differences are discussed in this
subsection.

From (\ref{dmcomplex}) we see that the phases $\phi$ and $\phi'$ do not
affect directly the running of the masses. However, they have an important
influence in the stability of a given choice of the matrix $U$, and
this in turn will influence the evolution of the masses. For example, in the
degenerate case $m_1\simeq m_2=m_0$, the initial choices $\phi-\phi'=0$ and
$\phi-\phi'=\pi$ (which correspond to $\tilde{m}_1\simeq \pm \tilde{m}_2$,
already analyzed in the real case) give identical $d m_i/dt$ at $t=0$.
However, in the first case $U$ changes abruptly, while in the second
it evolves smoothly, and this has an evident effect on the subsequent
running of $m_{1,2}$. The phase $\delta$, on the other hand, 
enters $U_{31}$ and $U_{32}$, and thus has a direct
effect on the size of $d m_i/dt$, although its importance is somewhat
reduced by the smallness of $\theta_2$. 

One important instance in which the stability of $U$ is crucial for the
viability of the scenario is the VO solution to the solar neutrino problem.
As discussed in the real case, in a scenario with total or partial 
degeneracy of neutrino masses,
the final $\Delta m_{21}^2$ has to be much
smaller than the typical size of the RG shift in the masses squared
$\sim 2 m_0^2\epsilon_\tau$. Hence, starting with $m_1\simeq m_2\gg m_3$ at
the 
high energy scale
$\Lambda$, to keep $\Delta m_{21}^2$ under control requires an exquisite
cancellation in the running of $m_1$ and $m_2$. From (\ref{dmcomplex}), this
cancellation is 
\be
\label{cancelcomplex}
|(|U'_{31}|^2-|U'_{32}|^2)|\simlt \frac{\Delta m_{sol}^2}
{4 m_0^2 \epsilon_\tau}\ .
\ee
In this condition, we have written $U'$ and not $U$, because 
if the degeneracy is good enough $U$ jumps to $U'$ almost immediately away
from $\Lambda$ and
remains close to it during the rest of the running down to $M_Z$. By
definition, $U'$ satisfies the stability condition
\be
\label{stablecomplex}
Re({U'}_{31}^*U'_{32})=0.
\ee
From (\ref{cancelcomplex}) we find that, to keep the solar mass splitting 
under control, the mixing angles (at low energy) must be correlated
according to 
\be
\label{cancelexact}  
\tan 2\theta_3\simeq \frac{\cos^2\theta_1 \sin^2\theta_2
-\sin^2\theta_1}{\sin\theta_2\sin 2\theta_1\cos\delta}\ .  
\ee 
This equation generalizes (\ref{U31U32}) to the complex case.
Eqs.~(\ref{stablecomplex}, \ref{cancelexact}) must be satisfied
simultaneously for the viability of the VO scenario in totally 
or partially degenerate scenarios. 

From eq.~(\ref{cancelexact}) we can
see that a non-zero $\delta$ will influence the correlations that must
occur. The effect is discussed in figure 1, where, for two different values
of $\sin^22\theta_1$ (the maximal one, 1, and the experimental lower bound,
0.82) we plot the region of the plane ($\sin^22\theta_2,\sin^22\theta_3$)
where the cancellation (\ref{cancelcomplex}) takes place. The different
strips\footnote{For a fixed value of $\sin^22\theta_2$ and
$\sin^22\theta_1$ there are in
principle 4 values of $\sin^22\theta_3$ satisfying (\ref{cancelexact}),
corresponding to the fact that the interchange $\cos\theta\rightarrow
\sin\theta$ leaves  $\sin^22\theta$ invariant. We do not plot the two
solutions with $s_2>c_2$, which would be in conflict with CHOOZ+SK data.
Note that, for $\sin^22\theta_1=1$ the two remaining solutions merge in
one. } correspond
to different values of $\cos\delta$, from 1 (darker color) to $0$
(lighter), in $1/4$ steps. We set the high energy scale $\Lambda$ at the
typical see-saw value $10^{12}\, {\rm GeV}$ and require $\Delta
m_{12}^2\leq
1.1\times 10^{-10}\, {\rm eV}^2$ with the initial neutrino mass
$m_0^2=\Delta m_{atm}^2$ $=5\times 10^{-4}\, {\mathrm eV}^2$. For much
smaller $\Lambda$ the strips in figure 1 get thicker [it is easier to
satisfy (\ref{cancelcomplex}) because the size of radiative corrections is
smaller]. Note that $\delta=\pi/2$ is special because, in such case, 
eq.~(\ref{cancelexact}) has in
principle two solutions: {\em a)} $\sin^22\theta_3=1$ and {\em b)}
$s_2^2=\tan^2\theta_1$.
The region satisfying (\ref{cancelcomplex}) will consist of two strips
centered
around those two lines.
In the case $\sin^22\theta_1=1$, the solution {\em b)} gives
$s_2^2=1$, which
is beyond experimental bounds and only solution {\em a)}, 
$\sin^22\theta_3=1$, remains
(the region around that line has zero width and cannot be seen in the
plot).
For $\sin^22\theta_3<1$, solution {\em b)} can have $s_2<c_2$ and we
include it in the figure.
The corresponding region is T-shaped like the one shown in figure 1,
lower plot.

On the other hand, the condition (\ref{stablecomplex}) for stability 
can be satisfied for any
choice of mixing angles and $\delta$ by adjusting $\phi-\phi'$ to the
apropriate value, which is given by
\be
\label{opcom}
\tan[(\phi-\phi')/2]=\frac{\sin 2\theta_3
(\sin^2\theta_1-\cos^2\theta_1\sin^2\theta_2)
-\sin2\theta_1\cos2\theta_3\sin\theta_2\cos\delta}{
\sin2\theta_1\sin\theta_2\sin\delta}\ .
\ee
This would be the complex version of the requirement of opposite signs 
for the two degenerate eigenvalues in the real case.
Using (\ref{cancelexact}), the previous result simplifies further to
\be
\tan[(\phi-\phi')/2]=-\frac{1}{\tan\delta\cos2\theta_3}.
\ee
Unlike the real case, to impose condition (\ref{opcom}) for generic values
of the parameters, seems hard to justify from some underlying symmetry.

\section{Conclusions}

Assuming  three flavours of light Majorana neutrinos, with no
reference to any particular scenario,  
we have derived the general renormalization group equations
(RGEs) for the physical  neutrino parameters: three masses, 
three mixing angles and three CP-violating phases [information
alternatively encoded in the RGE for the masses and the complex mixing
(MNS) matrix $U$]. This form of writing the RGEs represents an
advantageous alternative to using the RGE for ${\cal M}_\nu$. It
 avoids the proliferation of
unphysical parameters, which allows to keep track of the
physics in a more efficient way. It also permits to 
 appreciate interesting features, \eg\ presence
of stable (pseudo infrared fixed-point) directions for mixing angles and
phases, which are not  consequence of a particular scenario. 

We have then particularized the RGEs for relevant scenarios. Namely,
when the effective theory below $\Lambda$ (the scale at which the
neutrino mass operator is generated), is given by the SM or MSSM; or
when  ${\cal M}_\nu$ is generated by a see-saw mechanism. In the first
two scenarios we have analyzed in detail the physical implications of
the RGEs, separating the case where $U$ is real (\ie\ no CP
phases) and the general complex case.

For the real case, we have noticed that if one starts with two
masses suficiently degenerate (in absolute value and sign), say
$m_i\simeq m_j$, the $U$ matrix is driven to a stable  (infrared
pseudo-fixed point) form, providing net predictions  for the mixing
angles. Whenever this happens the corresponding mass splitting,
$\Delta m^2_{ij}$, is entirely determined  at low energy by the 
RG running. This amounts to a very predictive scenario. Depending on 
what are the initial (quasi) degenerate neutrinos, the scenario 
can be realistic or not. In particular, starting with $m_1\simeq m_2$
and an initial $\Delta m^2_{32}\sim \Delta m^2_{atm}$, we finish 
at low energy with a small ``solar'' $\theta_3$ angle and a
$\Delta m^2_{21}$ splitting just of the right size for the 
SAMSW solution to the solar neutrino problem, which is certainly 
remarkable.
On the other hand, the radiative corrections to the mass splittings are
potentially dangerous for the VO scenario, which becomes unviable if 
$m_1^2\simeq m_2^2\gg m_3^2$, unless $m_1\simeq -m_2$, the ``solar'' 
$\theta_3$ angle is close to maximal and the common mass is below 
the range of cosmological relevance.

We have also shown that previous claims (realized in the 
two-flavour approximation) in the sense that the RG running could provide a
substantial enhancement of the atmospheric mixing, $\sin^2 2\theta_1$ 
cannot work in practice, since the mechanism leads to 
an unacceptably large ``solar'' splitting, $\Delta m^2_{21}$.
We have shown that, unfortunately, this is also the case in the 
more general 3-flavour scenario.

For the general complex case,  most of
the conclusions are similar, but there are interesting new effects. In
particular, the
previously considered ``radiative'' SAMSW
scenario requires  $\phi-\phi'\simeq 0$ (unless the ``solar'' $\theta_3$
angle is set by hand  at a small value from the beginning). On the other
hand, if $\phi-\phi'$ is different from zero and the neutrino spectrum has
a  two-fold degeneracy, the RGEs will
generically drive its value to $\pm \pi$. It is also worth-noticing
that the CP phase  $\delta$ can be driven to maximal values by RG
corrections.  On the other hand, if $\cos\delta\neq 0$, the
viability of the VO scenario when $m_1^2\simeq m_2^2$ requires, besides
the previous conditions, a delicate cancellation involving the three CP
phases, which makes the scenario more unnatural.

\section*{Addendum}

Shortly after the completion of this work, there appeared a paper by
P.H. Chankowski, W. Kr\'olikowski and S. Pokorski \cite{poko}
dealing with similar
subjects. More precisely, they work out the RGEs for mass eigenvalues
and mixing angles below the high energy scale $\Lambda$, assuming that
the effective theory below $\Lambda$ is the SM or MSSM and a real MNS
mixing matrix, $U$. Therefore, their work corresponds to the issues
studied here in sect.~3.
As far as we have checked, their results are in agreement
with ours. They also notice and stress the existence of stable
(infrared pseudo-fixed points) in the evolution of the mixings.

There is a point however in which we disagree. They work out the
2-flavour approximation, reaching an RGE for the (atmospheric) angle
$\theta_1$ which coincides with our equation (\ref{thetanuestra}) [or
equivalently  eq.~(\ref{sin2tnuestra})], but they argue that the
equation normally used in the literature, eq.~(\ref{Babutheta}), is
incorrect and inconsistent with (\ref{thetanuestra},
\ref{sin2tnuestra}), which is not the case (once
the difference between mass matrix entries and mass
eigenvalues is taken into account).

On the other hand they argue, correctly, that the usual claim that a
maximal angle is RG stable is not correct. Actually, we have shown 
that this is also the case in a 3-flavour scenario, giving
the more general conditions which would guarantee stability.

Finally, let us remark that in our paper we study also the general complex
case,  \ie\ when CP phases are present. In addition, in sect.~2 we work out
the RGEs for a completely general case (with no reference to \eg\ SM or
MSSM), showing explicitly the appearance of stable (infrared pseudo-fixed
points) forms for $U$.

\section*{Appendix}

We derive here the general form of the stable (infrared pseudo-fixed point)
 MNS matrix, $U$, in the presence of quasi-degenerate neutrino
masses. The generic RGE of the neutrino mass matrix is given by
($t=\log\mu$)
\be 
\frac{d {\cal M}_\nu}{dt}=-(\kappa_U
{\cal M}_\nu + {\cal M}_\nu P + P^T {\cal M}_\nu),
\ee
from which the RGEs for the mass eigenvalues $m_i$ and the mixing matrix
$U$ are obtained as
\be
\label{RGmassap}
\frac{d m_i}{dt}= - 2 m_i \hat P_{ii} - m_i Re(\kappa_U),
\ee
and
\be
\label{dUdt} 
\frac{dU}{dt}=UT .
\ee
$T$ is an anti-hermitian matrix given by
\bea
T_{ii}&\equiv&i \hat{Q}_{ii},\nonumber\vspace{.5cm}\\
T_{ij}&\equiv&\nabla_{ij} Re (\hat P_{ij})+i[\nabla_{ij}]^{-1} Im
(\hat P_{ij})+i\hat{Q}_{ij},\hspace{1cm} i\neq j,
\eea
with $\hat{P} \equiv \frac{1}{2}U^\dagger (P+P^\dagger) U$,
$\hat{Q} \equiv -\frac{i}{2}U^\dagger (P-P^\dagger) U$
and $\nabla_{ij}\equiv (m_i+m_j)/(m_i-m_j)$.

Let us suppose that two mass eigenvalues, say $m_1$ and $m_2$, 
are almost degenerate at the initial high-energy scale $\Lambda$. Then,
near the starting point of the running, 
$t \sim t_0$, $T$ is dominated by the 
(real) terms proportional to $\nabla_{12}$
\be 
T\sim \pmatrix{0 &
-Re(T_{21}) & 0\cr Re(T_{21}) & 0 & 0 \cr 0 & 0 & 0} .
\ee
Then, the formal solution to eq.~(\ref{dUdt}) is given by
\be 
\label{Uprima}
U'\equiv U(t) \sim U
{\mathrm exp}\left[\pmatrix{0 & \int_{t_0}^t Re(T_{21})dt & 0\cr
-\int_{t_0}^{t}
Re(T_{21})dt & 0 & 0 \cr 0 & 0 & 0} \right]=U R_{12}(\Gamma) ,
\ee
where $R_{12}(\Gamma)$ is an ordinary rotation in the 1--2 plane 
by an angle $\Gamma$
\be
R_{12}(\Gamma)=\pmatrix{c_{\Gamma} & s_{\Gamma} & 0\cr -s_{\Gamma} &
c_{\Gamma} & 0 \cr 0 & 0 & 1\cr} .
\ee
The value of $\Gamma$ will be such that
$d U /dt$ becomes  non-singular, \ie\ it renders $Re(T_{12})\simeq 0$.
Thus the stable matrix $U'$ satisfies
\be
\label{rep120}
Re(\hat P'_{12})=\frac{1}{2}Re(U'^\dagger (P+P^\dagger) U')_{12}=0\ ,
\ee
which we can solve for $\Gamma$ in terms of initial quantities obtaining 
\be 
\label{2Gamma}
\tan2{\Gamma}
= \frac{2 Re (\hat P_{12})}{\hat P_{22}-\hat P_{11}}.
\ee 
Therefore, using (\ref{Uprima}), we get 
unambiguously the stable $U'$ matrix in terms of the initial $U$.

One can further justify the condition (\ref{rep120}) by showing that the RG
evolution actually drives $Re(\hat{P}_{12})$
towards zero, as expected. The relevant RG when $\nabla_{12}$ dominates is
\be
\label{apr12}
\frac{d}{dt}Re(\hat{P}_{12})\simeq (\hat{P}_{11}-\hat{P}_{22})\nabla_{12}
Re(\hat{P}_{12}).
\ee
For a sufficiently long running interval one has [using (\ref{RGmassap})]
\be
\nabla_{12}\simeq \frac{1}{(\hat{P}_{11}-\hat{P}_{22})\log(\Lambda/\mu)},
\ee
and inserting this in (\ref{apr12}) we get $Re(\hat{P}_{12})\rightarrow 0$ 
in the infrared.

Let us consider now the only remaining  possibility, \ie\ 
that all the mass eigenvalues $m_1$, $m_2$, $m_3$ are
(almost) degenerate. Similarly to the previous case, for
$t \sim t_0$, $T$ is dominated by the 
(real) terms 
\be 
T\sim \pmatrix{0 & -Re(T_{21}) & -Re(T_{31})\cr Re(T_{21}) & 0 &
-Re(T_{32}) \cr Re(T_{31}) & Re(T_{32}) & 0} \ .
\ee
So, the solution to eq.~(\ref{dUdt}) is formally  given by
\be 
\label{Uprima2}
U'\equiv U(t) \sim U {\mathrm exp}\left[\pmatrix{0 &
\int_{t_0}^t Re(T_{21})dt & \int_{t_0}^t Re(T_{31})dt\cr -\int_{t_0}^{t}
Re(T_{21})dt & 0 & \int_{t_0}^t Re(T_{32})dt \cr -\int_{t_0}^t
Re(T_{31})dt & -\int_{t_0}^t Re(T_{32})dt & 0} \right] = UR , 
\ee
where $R$ is a general $3\times 3$ rotation, depending on three
angles, say $\Gamma_1$, $\Gamma_2$, $\Gamma_3$, 
\be 
\label{R}
R=\pmatrix{c_2c_3 & c_2s_3 & s_2\cr
-c_1s_3-s_1s_2c_3 & c_1c_3-s_1s_2s_3 & s_1c_2\cr
s_1s_3-c_1s_2c_3 & -s_1c_3-c_1s_2s_3 & c_1c_2\cr},
\ee
with $s_i=\sin \Gamma_i$, $c_i=\cos \Gamma_i$ (the resemblance 
with the 'CKM' matrix is only formal). $R$ is
determined, as in the case of twofold degeneracy considered before, by the
fact that $d U /dt$ becomes  non-singular, \ie\ $Re(T_{ij})\simeq 0$.
Thus
\be 
\label{gam1}
Re(\hat P'_{12})=\frac{1}{2}Re(U'^\dagger (P+P^\dagger)
U')_{12} =0, \ee 
\be \label{gam2}
Re(\hat P'_{31})=\frac{1}{2}Re(U'^\dagger
(P+P^\dagger) U')_{31}=0, \ee
\be \label{gam3}
Re(\hat P'_{32})=\frac{1}{2}Re(U'^\dagger (P+P^\dagger)
U')_{32}=0.
\ee
Substituting $U'=UR$, these equations determine completely the rotation
$R$, and therefore $U'$, in terms of $U$ and the initial parameters.
Another interesting remark is that if one has information on the mixing angles
in the stable matrix $U'$, the phases will be constrained, so as to satisfy
the set of equations (\ref{gam1}-\ref{gam3}). And viceversa, if the phases
are known, the mixing angles cannot take arbitrary values.

Again, one can show that in this case RG evolution drives $Re(\hat
P_{ij})\rightarrow 0$ for all $i,j$. The relevant RGE that shows this is
now
\bea
\frac{1}{2}\frac{d}{dt}\left\{[Re(\hat{P}_{12})]^2
+[Re(\hat{P}_{23})]^2+[Re(\hat{P}_{31})]^2\right\}&=&
(\hat{P}_{11}-\hat{P}_{22})\nabla_{12}[Re(\hat{P}_{12})]^2\vspace{0.1cm}
\nonumber\\
+(\hat{P}_{22}-\hat{P}_{33})\nabla_{23}[Re(\hat{P}_{23})]^2
&+&(\hat{P}_{33}-\hat{P}_{11})\nabla_{31}[Re(\hat{P}_{31})]^2 .
\eea
Using 
\be
\nabla_{ij}\simeq \frac{1}{(\hat{P}_{ii}-\hat{P}_{jj})\log(\Lambda/\mu)},
\ee
in the equation above, we get $Re(\hat{P}_{ij})\rightarrow 0$  in the
infrared for all $i,j$.
 
\section*{Acknowledgements}
A. I. thanks the Comunidad de Madrid (Spain) for a pre-doctoral grant. 


\end{document}